\documentclass[fleqn,usenatbib]{mnras}

\usepackage{mathptmx}

\usepackage[T1]{fontenc}
\usepackage{ae,aecompl}

\usepackage{graphicx}	
\usepackage{amsmath}	
\usepackage{amssymb}	
\usepackage{comment}
\usepackage{caption}
\usepackage{subcaption}

\usepackage{pdflscape}
\usepackage{hyperref}
\usepackage{microtype}
\usepackage{color}
\usepackage[utf8]{inputenc}

\newcommand{\HI}{H$\,${\sc i}}

\newcommand{\nhi}{$N_{\rm H\,I}$}

\newcommand{\lya}{Ly$\alpha$}

\graphicspath{{./}{figures/}}

\title[Feedback and Ly$\alpha$ Forest I]{\huge Can the Low Redshift Lyman Alpha Forest Constrain AGN Feedback Models?}
\author[Khaire et al.]
{
\parbox{\textwidth}{
Vikram Khaire,$^{1,2}$
Teng Hu,$^2$ Joseph F. Hennawi,$^{2, 3}$ Michael Walther,$^{4, 5}$ and \\
Frederick Davies$^{\,2,6}$ 
} 
\vspace*{10pt}\\
$^{1}$Indian Institute of Space Science \& Technology, Thiruvananthapuram, Kerala  695547, India\\
$^{2}$Physics Department, Broida Hall, University of California Santa Barbara, Santa Barbara, CA 93106-9530, USA\\
$^{3}$Leiden Observatory, Leiden University, PO Box 9513, NL-2300 RA Leiden, the Netherlands\\
$^{4}$University Observatory, Faculty of Physics, Ludwig-Maximilians-Universit\"{a}t, Scheinerstr. 1, 81677 M\"{u}nchen, Germany\\
$^{5}$Excellence Cluster ORIGINS, Boltzmannstrasse 2, D-85748 Garching, Germany\\
$^{6}$Max-Planck-Institut für Astronomie, Königstuhl 17, 69117 Heidelberg, Germany
}   
\pubyear{2023}
\begin{document}
\label{firstpage}
\pagerange{\pageref{firstpage}--\pageref{lastpage}}
\maketitle





\begin{abstract}
We investigate the potential of low-redshift Lyman alpha (Ly$\alpha$) 
forest for constraining active galactic nuclei (AGN) feedback models 
by analyzing the Illustris and IllustrisTNG 
simulation at $z=0.1$. 
These simulations are ideal for studying the impact of AGN feedback on 
the intergalactic medium (IGM) as they share initial conditions with
significant differences in the feedback prescriptions. 
Both simulations reveal that the IGM is significantly impacted by AGN feedback. 
Specifically, feedback is stronger in Illustris and results in reducing cool baryon 
fraction to 23\% relative to 39\% in IllustrisTNG. 
However, when comparing various statistics of Ly$\alpha$ forest such as
2D and marginalized distributions of Doppler widths and H~{\sc i} column 
density, line density, and flux power spectrum 
with real data, 
we find that most of these statistics are largely insensitive to the 
differences in feedback models. This lack of sensitivity arises because of the fundamental
degeneracy between the fraction of cool baryons and the H~{\sc i} photoionization rate
($\Gamma_{\rm HI}$)
as their product determines the optical depth of the Ly$\alpha$ forest. 
Since the $\Gamma_{\rm HI}$ cannot be precisely predicted from 
first principles, it needs to be treated as a nuisance 
parameter adjusted to match the observed Ly$\alpha$ line density.
After adjusting $\Gamma_{\rm HI}$, 
the distinctions in the considered statistics essentially fade away.
Only the Ly$\alpha$ flux power spectrum at small spatial scales exhibits potentially 
observable differences, although this may be specific to the relatively extreme 
feedback model employed in Illustris. 
Without independent constraints on either $\Gamma_{\rm HI}$ or cool baryon fraction,
constraining AGN feedback with low-redshift Ly$\alpha$ forest will be very challenging. 

\end{abstract}

\section{Introduction} \label{sec:intro}
One of the major unsolved problems in galaxy formation is the physical mechanism that 
quenches star formation in massive galaxies, resulting in the observed dichotomy between 
blue star-forming and red-and-dead elliptical galaxies. In order to reproduce this, 
cosmological simulations often include AGN feedback, which extracts energy from central 
supermassive black holes and drives powerful galactic outflows that suppress star 
formation in massive galaxies at late times. The mechanisms by which AGN feedback occurs, 
such as the episodic quasar mode triggered by major merger events \citep{Springel05, 
Hopkins08} or the continuous radio mode \citep{Croton06, Sijacki07}, are not yet fully 
understood and are a subject of ongoing debate. 
The multitude of modeling prescriptions used in 
recent simulations, such as Horizon-AGN \citep{Horizon_agn}, EAGLE \citep{Eagle}, 
Illustris \citep{Illustris}, IllustrisTNG \citep{IllustrisTNG} and  SIMBA \citep{Simba}, 
demonstrate the challenges of understanding this process and the importance 
of constraining it with 
observations.

In these simulations, AGN feedback drives powerful outflows 
that suppress star formation in massive 
galaxies and therefore can alter the physical state of gas in the circumgalactic 
and intergalactic media (CGM and IGM) surrounding these galaxies. 
This process, in addition to accretion shocks in the halos of galaxies,
creates a bubble of tenuous, warm-hot gas ($T \sim 10^{5-7}$ K, 
$n_{\rm H} \leq 10^{-4}$ cm$^{-3}$) around halos up to several Mpc in scale, 
known as the warm-hot IGM (WHIM). The amount of matter in the WHIM phase is related 
to the `missing baryon problem', which refers to the deficit of baryons 
in the present-day universe compared to the amount synthesized in the Big Bang 
\citep[e.g.][]{Persic92, Fukugita98}. Approximately 30-40\% of the 
total baryons are currently unaccounted for at $z < 1$ 
\citep[see e.g][]{Lehner07, Prochaska09, Shull12_baryons}, 
compared to the precisely determined abundance from the 
cosmic microwave background at $z \sim 1100$ and big bang nucleosynthesis \citep{Planck18} . 
These missing baryons are difficult to detect because the diffuse gas in the WHIM phase is 
notoriously challenging to observe. While there have been various claims of detection 
through X-ray emission \citep{Eckret15}, stacking the feature arising from 
the Sunyaev-Zeldovich effect \citep{Tanimura19, Graff19}, X-ray absorption lines 
\citep[e.g. O$\,${\sc vii} and O$\,${\sc viii};][but see  
\citealt{Johnson19}]{Nicastro19}, and countless Hubble Space Telescope (HST) UV
absorption spectroscopy studies modeling both  metals
\citep[O$\,${\sc vi} and Ne$\,${\sc viii};][]{Tripp08, Savage14, Werk2016_ovi,
Hussain17,Burchett19} and broad Lyman-$\alpha$ (\lya~hereafter) absorbers
\citep[e.g.][]{Savage11BLA, Narayanan12, Tejos16}, robust quantitative constraints
on the fraction of cosmic baryons in this WHIM phase have proven elusive.

Theoretically, the missing baryons can be tracked in hydrodynamic simulations 
\citep[e.g][]{Dave01_whim, Martizzi19_tng_whim, Tuominen21}, which predict that 
even in the absence of AGN feedback, accretion shocks in the halos of galaxies resulting from
structure formation alone implies that 30-50\% of baryons reside in 
filaments and halos at the nodes of the cosmic web, 
shock heated to $T \sim 10^{5-7}$ K by $z < 1$ \citep{Cen99, Cen06, Dave01}. 
The total amount of diffuse gas in the WHIM phase is a combination of the 
contribution from structure formation and the consequences 
of energetic feedback events \citep[see e.g][]{Tornatore10}. 

At $z > 2$, the vast majority of baryons are known to reside primarily in a 
$\sim 10^4$ K, photoionized phase probed by the \lya~forest 
\citep[$>80\%$;][]{Prochaska09}, and  cosmological simulations suggest that feedback 
from galaxies has a negligible impact on the IGM at these redshifts 
\citep{Kollmeier06, Viel13feedback, Chabanier20}. This is also 
supported by the percent level agreement between the statistical properties of the 
\lya~forest and simulations which do not include feedback 
\citep{Viel06, Lee15, Walther19}. However, whether AGN feedback
sufficiently
alters the structure of the \lya~forest at later times $z < 1$ is a 
fundamental open question.

To address this question, we compare the results of two state-of-the-art 
galaxy formation simulations: the original Illustris simulation \citep{Illustris} and its 
successor IllustrisTNG \citep{IllustrisTNG}. These simulations are well-suited for 
comparison because they share the same underlying code, initial conditions, and 
similar cosmological parameters. The main difference between them is in the implementation of 
galaxy formation feedback, specifically the persistent radio-mode AGN feedback. 
Illustris used strong radio-mode thermal feedback, while IllustrisTNG used 
mild kinetic radio-mode feedback, resulting in significant differences in the 
distribution of baryons around massive halos and in the intergalactic medium (IGM). 
In this work, we focus on the \lya~forest at a $z=0.1$ to investigate the impact of 
AGN feedback on the IGM.

We use high signal-to-noise quasar spectra from the low-redshift IGM survey of 
\citet{Danforth14} to generate forward-modeled
\lya~forest spectra from Illustris and IllustrisTNG simulations. 
We fit the \lya~absorption lines in these simulated spectra and the observed data using an 
automated Voigt profile fitting code. We use the results of it to fix the hydrogen 
photoionization rate ($\Gamma_{\rm HI}$) in both simulations in order to match the 
absorption line density ($\rm {\rm dN/dz}$) observed in the data. We show that, 
fixing $\Gamma_{\rm HI}$ in this way
is crucial for the analysis since it removes the degeneracy between the $\rm {\rm 
dN/dz}$ and $\Gamma_{\rm HI}$. We then examine various statistics, including the 
line-of-sight flux power spectrum,  the joint distribution of line widths ($b$) 
and neutral hydrogen (\HI) column densities ($N_{\rm HI}$), 
as well as their respective marginalized distributions, and \HI~column density distribution
function to assess the impact of AGN feedback on the IGM. Our results suggest 
that feedback has a non-negligible effect on the IGM, but the currently available data do 
not have sufficient signal-to-noise ratio or resolution to clearly detect it.

However, differences in the gas distribution around halos of massive 
galaxies between Illustris and IllustrisTNG are significant. In a forthcoming 
follow-up paper \citep[referred to as Paper II,][under review]{Khaire23}, the analysis is 
repeated exclusively around massive halos to investigate the effects of feedback.

The paper is structured as follows. In Section~\ref{sec.all_sims}, we describe the 
simulations used in the study, including their differences in the implementation of 
feedback and the resulting differences in the temperature and density of the gas in the 
IGM. In Section~\ref{sec.method_igm}, we outline the procedure for generating 
synthetic \lya~forest 
spectra from the simulations, generating forward models, and 
performing Voigt profile fitting. In Section~\ref{sec.gamma}, we discuss the 
importance and method for tuning the $\Gamma_{\rm HI}$. In Section~\ref{sec.igm}, 
we examine the IGM in both simulations using various statistics and discuss our results. 
The implications of our findings are discussed in Section~\ref{sec.disc}, 
and a summary is provided in Section~\ref{sec.summary}.

\begin{figure*}
\includegraphics[width=0.99\textwidth,keepaspectratio]{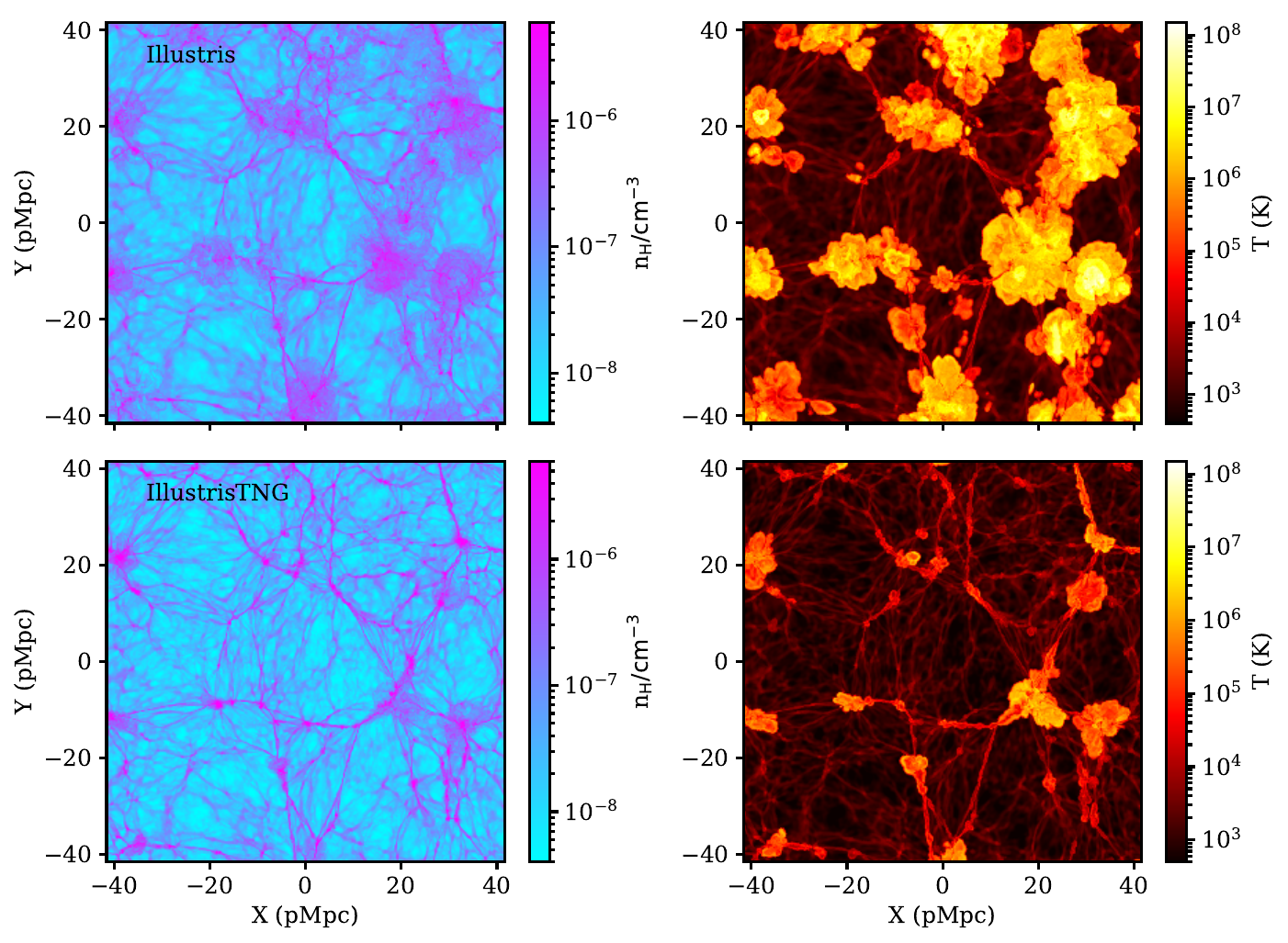}
\caption{The figure compares the density $n_{\text{H}}$ (left) and temperature $T$ (right)
of gas at the same location in Illustris (top) and IllustrisTNG (bottom) simulations at 
$z=0.1$. Each panel shows a 2D slice (size of $\sim 80$ pMpc) of the simulation with a 
single cell width of $\sim 60$ ckpc. 
The distribution of $n_{\text{H}}$ and $T$ is clearly 
different in the two simulations, particularly around the nodes in the cosmic web where 
halos reside. This difference is primarily due to the implementation of 
the radio-mode AGN feedback in both simulations; 
the explosive feedback in Illustris leads to the displacement of 
hot gas to a larger volume compared to IllustrisTNG.}
\label{fig.boxplot}
\end{figure*}
\section{Simulations}\label{sec.all_sims}
To investigate the impact of AGN feedback on the distribution of gas around galaxies and 
in the IGM, we use two state-of-the-art galaxy formation 
simulations: Illustris \citep{Vogelsberger14, Illustris} and IllustrisTNG 
\citep{IllustrisTNG, Pillepich18}. Both simulations have the same initial conditions, 
allowing us to compare the \lya~absorption along the same sightlines and study the 
gas around the same dark matter halos (as described in Paper II).

The main difference between the two simulations is in their feedback prescriptions 
(see Section~\ref{sec.sims_ill}).
While the cosmological parameters are also slightly different, we expect these 
small changes to have a negligible effect on the gas distribution compared to the 
impact of the feedback itself. Therefore, we ignore these differences in our analysis. 
However, we do use the same cosmological parameters that were used for running each 
simulation (see Table~\ref{tab.cosmo}) in our analysis of each simulation.

It is well known that the impact of feedback is more important and evident 
at the lowest redshifts. Therefore, we select simulation snapshots at 
$z=0.1$ for both Illustris and IllustrisTNG for the main analysis. 
The choice of this redshift is also based on the availability of 
observed low-$z$ \lya~forest data 
from Cosmic Origin Spectrograph (COS) on board HST 
(and the number of the quasar-luminous red galaxy
pairs that we consider in Paper II). For Illustris\footnote{Illustris:
\href{https://www.illustris-project.org}{https://www.illustris-project.org}}
and IllustrisTNG\footnote{IllustrisTNG:
\href{https://www.tng-project.org}{https://www.tng-project.org}}
we use the publicly available simulation outputs with a box size of $75$ cMpc/h
and 1820$^3$ baryon and dark matter particles.
This choice of resolution and box size out of all the available runs 
by Illustris collaboration was based on two criteria:
i) we want to get the mock \lya~forest spectrum to have an average pixel 
scale (which is 4.1 km s$^{-1}$ when we deposit the simulation outputs onto 1820$^3$ 
Cartesian grids as mentioned later)
better than the sampling of low-$z$ \lya~forest HST COS data (with resolution 15-20 km s$^{-1}$), but also 
ii) the abundance of high mass
halos in the simulation volume should be sufficiently large to probe the statistical 
properties of gas around them (as studied in Paper II). 

In order to analyze the simulation data, we interpolate the quantities such as 
temperature, density, and velocity from the Voronoi mesh outputs of both Illustris and 
IllustrisTNG snapshots onto 1820$^3$ Cartesian grids. 
We use a Gaussian kernel with a size equal to $2.5$ times the radius of each 
Voronoi cell, assuming each cell is spherical, for the smoothing process. 
It should be noted that this is simply an approximation and the 
value of $2.5$ is arbitrary, however, this is a commonly used approach for Illustris and 
IllustrisTNG simulations (as confirmed through private communication with D. Nelson).

In the following sections, we will outline the specific details of the simulations and 
the differences in their feedback prescriptions relevant to the analysis presented here 
as well as in Paper II.

\subsection{The Illustris and IllustrisTNG Simulations}
\label{sec.sims_ill}
Illustris and IllustrisTNG are both cosmological hydrodynamic simulations that 
use the {\sc arepo} code \citep{Springel10}  and the tree-PM scheme  \citep{Bode00} to 
solve the Euler equations on a quasi-Lagrangian moving Voronoi mesh and calculate 
gravitational forces, respectively. Both simulations include various processes related to 
galaxy formation, such as star formation, stellar and AGN feedback, galactic winds, and 
chemical enrichment \citep[e.g][]{TNG_metals}. They also include photoionization heating and cooling, including 
metals, using the same UV background model of \cite{FG09}, 
as well as other relevant processes needed to 
model the \lya~forest, such as collisional ionization and inverse Compton cooling off the 
cosmic microwave background. However, one key difference is that IllustrisTNG includes 
ideal magnetohydrodynamic calculations, which are not present in Illustris.
\begin{figure*}
\includegraphics[width=0.98\textwidth,keepaspectratio]{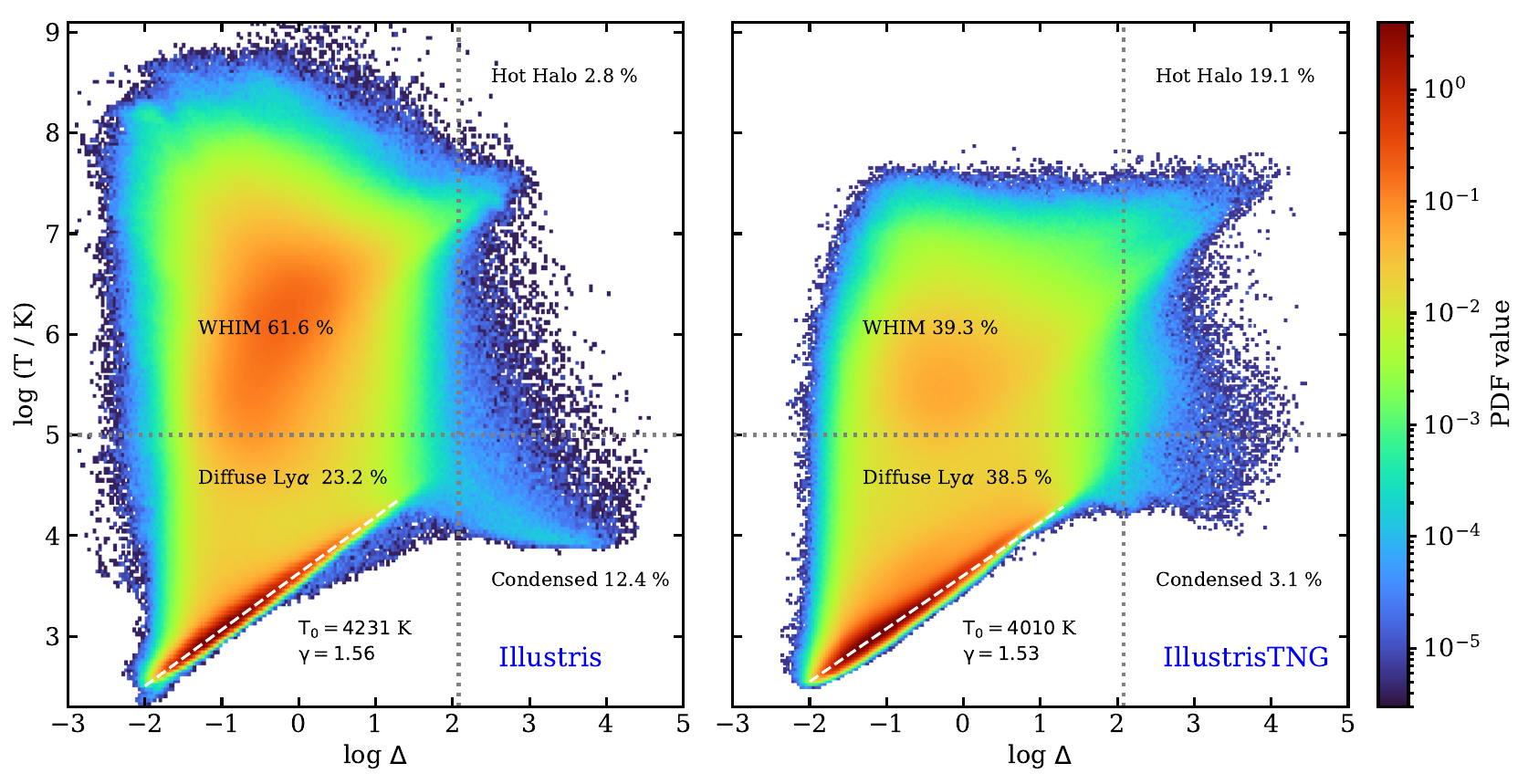}
\caption{The figure shows the density-temperature distribution of the IGM gas in both  
Illustris (left panel) and IllustrisTNG (right panel) simulations. The white dashed lines 
represent the power-law fit to the equation of state of the IGM, 
$T(\Delta) = T_0 \Delta ^{\gamma -1}$, with the best fit values of $T_0$ and 
$\gamma$ indicated in the legends. Dotted lines at $\Delta = 120$ and $T = 10^5$ K 
divide the phase diagram into four quadrants, corresponding to different gas phases: 
diffuse \lya~(cool baryons), WHIM, hot halo gas, and condensed gas. The mass fraction in each phase is 
also indicated in the legends (see also Table~\ref{tab.gas_phases}).
}
\label{fig.T-Delta}
\end{figure*}

The parameters governing the feedback prescriptions in galaxy formation simulations are 
often adjusted to match various observations, and Illustris and IllustrisTNG are no 
exceptions. The observations used for guiding these adjustments in 
Illustris and IllustrisTNG include the star formation rate density, 
stellar mass function \citep{TNG_Stars}, galaxy sizes and color distribution \citep{TNG_galaxy_color}, clustering of galaxies \citep{TNG_clustering}, and 
total gas fraction within a radius of halos $R_{500c}$\footnote{$R_{500c}$ is the 
the radius of a sphere around halos enclosing the mass density 500 times the 
critical matter density.}\citep[e.g][]{IllustrisTNG, Genel18_galaxy_sizes}. 
Many of the galaxy formation models and the feedback parameters are different for 
these simulations \citep[see][]{Pillepich18}. 
In fact, the new feedback parameters and prescriptions were implemented in
the IllustrisTNG simulation in order to overcome a few shortcomings of the Illustris 
simulation \citep[refer to Figure 4 of ][]{Pillepich18} 
which include the underestimation of the gas fraction present within $R_{500c}$ of massive 
halos in Illustris. 
A comprehensive summary of the changes made 
to the feedback parameters and prescriptions can be found in Table 1 of 
\citet{Pillepich18}, with relevant changes briefly mentioned below.

The feedback mechanisms can be divided into two categories: stellar feedback, which is 
important for low-mass galaxies, and AGN  feedback, which is important for high-mass 
galaxies. In both simulations, AGN feedback is implemented in two modes based on the rate 
at which gas is accreted onto the central supermassive black hole: the 
quasar-mode \citep{Springel05, Hopkins08, Debuhr11}, when 
the black hole is in a high-accretion state and the 
radio-mode \citep{Croton06, Bower06, Sijacki07}, 
when it is in a low-accretion state. While both simulations use continuous thermal 
feedback in the quasar-mode, they differ in their implementation of radio-mode feedback. 
Illustris uses a bubble  model \citep{Sijacki07} in which a large amount of feedback
energy is gathered over some time and then released explosively, driving 
the gas out of massive halos into the IGM. In contrast, IllustrisTNG uses 
kinetic wind feedback \citep{TNG_outflows}, in which momentum is stochastically imparted to nearby cells from 
the low-accretion state black hole \citep[see][]{IllustrisTNG}. 
This latter approach has been shown to be effective in reproducing 
star formation rates, stellar mass functions, and galaxy color distributions
for halo masses of  $\sim 10 ^{12}$ M$_{\rm \odot}$  without 
expelling too much gas and becoming inconsistent with total gas fractions within 
$R_{500c}$ \citep{Dylan18_color, Pillepich18}.

The Illustris and IllustrisTNG simulations offer an ideal opportunity to study the impact 
of AGN feedback on the gas distribution around galaxy halos and in the IGM, due to their 
differences in the radio-mode AGN feedback prescriptions as described above. These 
publicly available large simulations have been run with the same initial conditions, 
allowing us to identify the same locations and compare the differences in the density and 
temperature distribution of the gas. One such comparison is illustrated in 
Fig.~\ref{fig.boxplot} where we show slices of the simulation boxes for the 
distribution of hydrogen density $n_{\rm H}$ and temperature $T$. The slice has a 
thickness of one grid cell (i.e., $\sim 60$ ckpc). The figure demonstrates that 
the distribution of $n_{\rm H}$ and $T$ are significantly different, particularly around 
the nodes of the cosmic web where halos reside. This is mainly due to the different 
implementations of radio-mode AGN feedback in the two simulations: the explosive feedback 
in Illustris results in the displacement of hot gas surrounding halos to a larger volume 
compared to IllustrisTNG. There are only a few regions, far from Illustris halos, where 
the density and temperature profiles are similar to those in IllustrisTNG (e.g. at the 
centre of the slice shown in Fig.~\ref{fig.boxplot}).

The differences in the temperature and density at the nodes of the cosmic web in 
Fig.~\ref{fig.boxplot} are also  reflected in the temperature-density 
phase diagram shown in  Fig.~\ref{fig.T-Delta}. 
It is created by randomly selecting a large number of pixels 
from the grids of temperature and density data (in terms of overdensity $\Delta$)
and plotting a normalized 2D histogram of 
the mass-weighted temperature and density values. 
A comparison of the two simulations reveals that Illustris has more gas at 
high temperatures ($> 10^5$ K), as expected based on the distribution shown in 
Fig.~\ref{fig.boxplot}. To quantify the gas mass in different phases, 
the phase diagram is divided into four quadrants using the lines $T = 10^5$ K and 
$\Delta = 120$, following the definition given in \citet{Dave10}. 
These quadrants represent four standard gas phases: diffuse Ly$\alpha$
gas or cool baryons ($T < 10^5$ K and $\Delta < 120$), 
warm-hot intergalactic medium or WHIM ($T > 10^5$ K and $\Delta < 120$), 
hot halo gas ($T > 10^5$ K and $\Delta > 120$), 
and condensed gas ($T < 10^5$ K and $\Delta > 120$). 
The condensed gas particles are those that eventually form stars in the galaxy 
formation simulation. The mass fractions in these four phases for 
Illustris and IllustrisTNG are shown in Fig.~\ref{fig.T-Delta} and listed in Table 
\ref{tab.gas_phases}. Compared to IllustrisTNG, Illustris has a lower fraction of gas in 
the diffuse \lya~phase (23.3\% versus 38.5\%) and a higher fraction in the WHIM 
(61.6\% versus 39.3\%), indicating that the more explosive feedback in Illustris 
effectively heats low-density gas ($\Delta < 120$) to high temperatures ($T> 10^5$ K)
reducing the fraction of diffuse \lya~gas and increasing the WHIM.

\begin{table}
\centering
\caption{Parameters of cosmology and $T-\Delta$ relation (at $z=0.1$)}
\begin{tabular}{cccc}
\hline
Parameters & Illustris & IllustrisTNG  \\
\hline
${\Omega_m}$       &  0.2726 &  0.3089  \\
$\Omega_{\Lambda}$ &  0.7274 &  0.6911  \\
$\Omega_{b}$       &  0.0456 &  0.0486  \\ 
$h$                &  0.704  &  0.6774  \\
$\sigma_{8}$       &  0.809  &  0.8159  \\
$n_s$              &  0.963  &   0.97   \\
\hline
$T_0$               &  4231 K  &  4010 K  \\
$\gamma$            &  1.56   &    1.53  \\
 \hline
\end{tabular}
\label{tab.cosmo}
\end{table}
\begin{table}
\centering
\caption{Gas mass fraction in different phases at $z=0.1$}
\begin{tabular}{ccccc}
\hline
Simulations & Diffuse & WHIM &  Hot halo & Condensed \\
&\lya & &\\
\hline
Illustris & 0.232 & 0.616 & 0.028 & 0.124 \\
IllustrisTNG & 0.385 & 0.393 & 0.191 & 0.031 \\
 \hline
\end{tabular}
\label{tab.gas_phases}
\end{table}

The temperature-density phase diagram in the post-\HI-reionization era ($z<6$) often 
exhibits a tight power-law relation, with most of the gas follows
$T(\Delta) = T_0 \Delta^{\gamma - 1}$, where $\rm T_0$ is the temperature at mean density 
(i.e at $\Delta =1$) and $\gamma$ is an adiabatic index \citep{Hui97, McQuinn16}. 
At high-$z$, about 80\% of the gas follows this power-law relation, 
but at low-$z$ only a small fraction does due to the presence of a significant 
amount of hot gas in the WHIM phase. Additionally, the gas tracing the power-law 
relation is more dispersed at low-$z$, as shown by the puffier high-density 
regions at low-$T$ and $\Delta$ values in Fig.~\ref{fig.T-Delta}. 
Therefore, it is not straightforward to fit a $T-\Delta$ power-law at low-$z$. 
To overcome this issue, we use the method presented in \citet{Teng22}.
We first select gas with $T < 10^5$ K and a range of gas overdensity 
$-1.5 < \log \Delta < 1$, then divide this region into 15 equispaced $\log \Delta$ bins. 
For the $i^{\rm th}$ bin, we record $\log \Delta_i$ as the median overdensity value and 
calculate the temperature $\log T_i$ at which the marginal distribution 
$P(\log T | \log \Delta)$ peaks, as well as its ``error" $\sigma_{T_i}$, 
which we define to be of half of the temperature range
that sets the 16\% to 84\% percentile ranges. 
We then fit the $T-\Delta$ power-law relation to  
($\log T_i$, $\log \Delta_i$) using these errors 
by minimizing the least-square difference.
We found that the ${T- \Delta}$ relations in both the Illustris and IllustrisTNG 
simulations are nearly identical, with $\rm T_0$ values of $4231$ K and $4010$ K 
and $\gamma$ values of $1.56$ and $1.53$, respectively 
(also noted in Table~\ref{tab.cosmo}). The power-law fits are illustrated in 
Fig.~\ref{fig.T-Delta} by the white dashed lines on the ${T- \Delta}$ phase 
diagrams. The agreement of the parameters governing the power-laws between the two 
simulations is perhaps not surprising, since the temperature of the gas following this 
power-law relation is governed by the photoionization heating from the UV background 
which is the same in both 
simulations. But this does indicate that while feedback processes alter the 
structure of the phase diagram, the ${T- \Delta}$ relation is nevertheless 
unchanged. Therefore measuring the ${T- \Delta}$ relation at low-$z$ may not 
provide any interesting clues regarding the feedback \citep[see e.g,][]{Hu23_whim}. 

Despite the similarity in the temperature-density power-law relation, the mass fractions 
of different gas phases are significantly different in both simulations 
\citep[see also][]{Viel13feedback}. 
These differences in the ambient IGM (as seen in Fig.~\ref{fig.T-Delta}) 
and in the temperature and density of gas around the halos (as shown in 
Fig.~\ref{fig.boxplot}) hints that we can use the \lya~forest to distinguish 
between different feedback models. 
The \lya~forest is sensitive to changes in hydrogen density and temperature, making it 
a useful tool for understanding the impact of different feedback approaches on the gas 
surrounding galaxies and in the ambient IGM. 
We will explore this further in the following sections.

\section{Methods to simulate Lyman-$\alpha$ forest}\label{sec.method_igm}
In this section, we will outline the steps we took to generate synthetic 
\lya~forest spectra and create mock HST-COS data using forward models. 
We will also discuss our use of automated Voigt profile fitting following \citet{Hiss18}.

\subsection{Generating Synthetic \lya~Forest Spectra}\label{sec.moc}

In order to generate synthetic \lya~forest spectra, we draw a large number 
of sightlines (10$^5$)  through both the Illustris and IllustrisTNG simulations. These 
sightlines are parallel to the (arbitrarily chosen) $z-$axis and extend from one side of 
the simulation box to the other. Starting points of these sightlines are chosen at random 
in the $x-y$ plane. Along these sightlines, we record the values of the 
temperature $T$, overdensity $\Delta$, and the peculiar velocity component parallel to the line 
of sight $v_z$. In addition to these quantities, we also need to calculate the neutral 
hydrogen fractions to create the synthetic \lya~forest spectra. We do this by 
solving for the neutral fraction in ionization equilibrium, taking into account both 
collisional ionization and photoionization. At every cell along a sightline, we use the 
values of $T$, $\Delta$, and $\Gamma_{\rm HI}$ to 
calculate the neutral fractions. We assume a constant value of $\Gamma_{\rm HI}$ for the 
entire simulation box. The specific values of $\Gamma_{\rm HI}$ used for Illustris and 
IllustrisTNG simulations are discussed in Section~\ref{sec.gamma}.

\begin{figure*}
\includegraphics[width=0.98\textwidth,keepaspectratio]{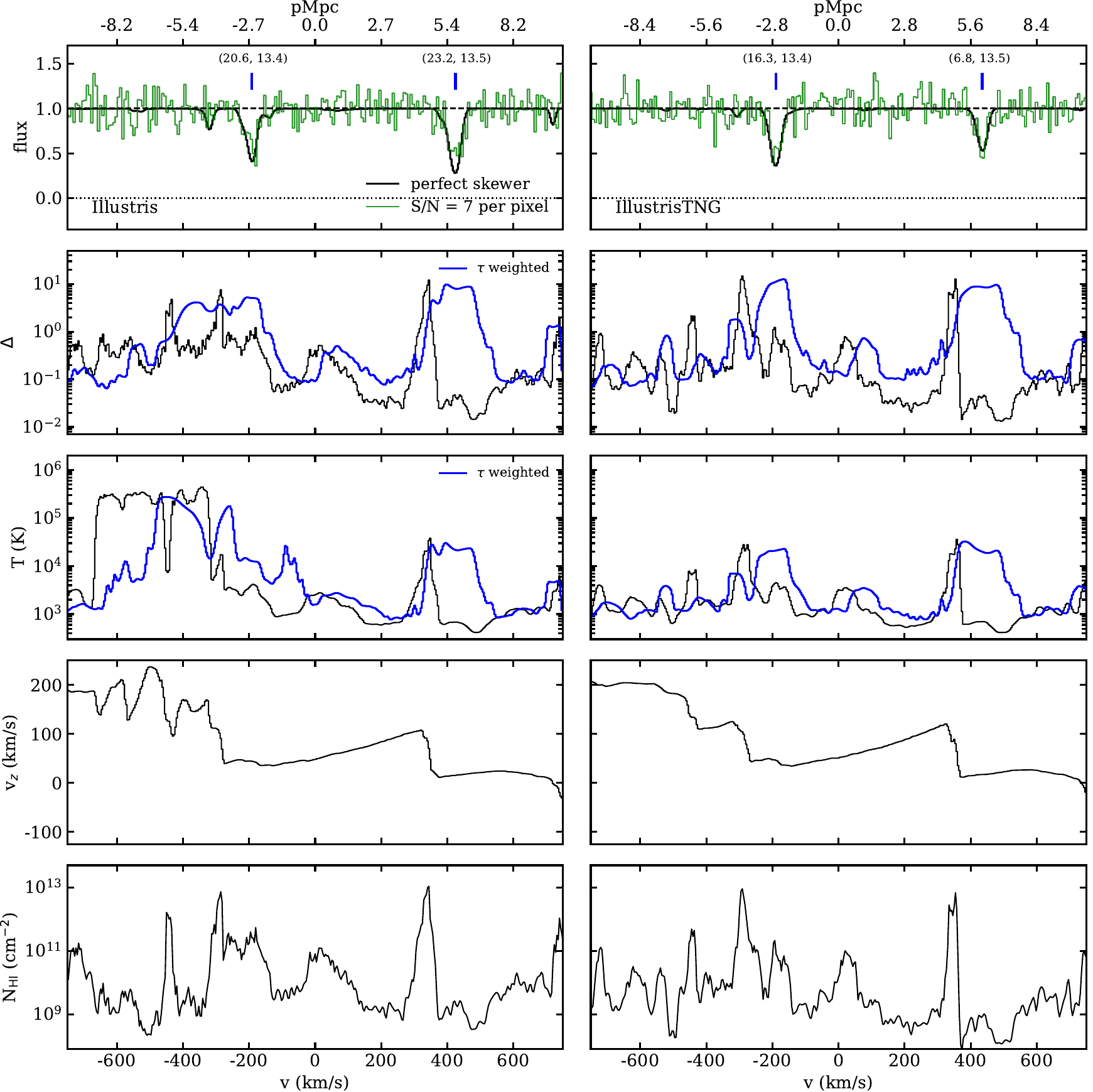}
\caption{Figure shows an example of a synthetic Ly$\alpha$ forest skewer 
(a chunk of 1500 km s$^{-1}$) from both Illustris (left) and IllustrisTNG (right) 
simulations. 
Both skewers are drawn along the same coordinates in both simulations and are 
shown from top to bottom with the normalized flux, overdensities ($\Delta$), temperature, 
line of sight velocity ($v_z$) and $N_{\rm HI}$ per pixel of the simulation
plotted on the y-axis. The green histograms on the top panels show the forward modeled 
skewers with a signal-to-noise ratio (S/N) of 7 per pixel (6 km s$^{-1}$) of the
forward model spectra convolved with the 
HST-COS G130M LSF. The blue ticks indicate the \lya~absorption lines identified by 
VPfit, with the fitted Doppler widths ($b$ in km s$^{-1}$) and column densities 
($\log N_{\rm HI}$ in cm$^{-2}$) shown in the legend. 
The blue dashed lines in the second and third panels from the top show the optical depth-
weighted overdensities and temperatures, respectively. The different sizes of the skewers 
in the left and right panels (as noted on the top x-axis) are a result of the different 
values of the $H_{\rm 0}$ used in the simulations.}
\label{fig.skewers}
\end{figure*}

To compute the ionization equilibrium, we need the electron density, which has a 
contribution (up to 16\%) from the ionization of helium gas. We approximate the 
contribution of helium to the electron density by assuming that the 
helium is doubly ionized i.e. the fractions of He~{\sc i} and He~{\sc ii} are zero. 
This approximation is justified because the redshift of the 
simulations is $z=0.1$, which is 10 billion years after the completion of helium 
reionization \citep[$z \sim 3$;][]{McQuinn09, Shull10, Worseck11, Khaire17sed}. 
Moreover, this simplifies 
calculations by removing two extra free parameters, the He~{\sc i} and He~{\sc ii} 
photoionization rates, from the analysis. Our ionization equilibrium calculations use 
updated cross-sections and recombination rates from \citet{Lukic15}. In addition, we also 
use the self-shielding prescription presented in \citet{Rahmati13} for dense cells.

After determining the neutral fraction of hydrogen, we proceed to generate the simulated 
optical depth $\tau$, for each cell along the line-of-sight ($z$-axis) by 
using the $T$ and the velocity component ($v_z$). This is achieved by summing the 
contributions from all real-space cells to the redshift 
space optical depth, considering the full Voigt profile resulting from each real space
cell following Voigt profile approximation given in \cite{Tepper06}. 
The resulting continuum-normalized flux $F$
of the Ly$\alpha$ forest is then obtained by evaluating $F = e^{-\tau}$ 
along each line of sight. These simulated spectra are referred to as perfect \lya~forest 
skewers and have a pixel scale of $4.12$ km s$^{-1}$ set by the $1820^3$ 
cells used in both simulations. This pixels scale is $\sim 4$ times smaller than the 
resolution of HST-COS spectra, which have FWHM$\sim 15-20$ km s$^{-1}$.

\subsection{Forward models for IGM and Voigt profile fitting}\label{sec.fwd_models}

To study the IGM in the simulations, we forward model the perfect 
\lya~forest skewers to generate mock HST-COS  \lya~forest spectra 
with S/N ratio and resolution similar to those presented in \citet{Danforth14}.
We use the spectra from \citet{Danforth14} because it contains the highest S/N low redshift 
($z<0.5$) \lya~forest data to date. 
We mask the spectral regions having metal lines from Milky Way interstellar medium (ISM)
and intervening IGM using the line identification provided by \citet{Danforth14}.
For this, we follow the same procedures outlined in \citet{Khaire19}. 
We forward model the \lya~forest in a redshift bin at $0.06<z<0.16$ which is probed 
by 34 high-quality (S/N per pixel > 10) quasar spectra with a median 
redshift of the \lya-forest $z=0.1$. 
To generate these forward models, we randomly stitch together the perfect skewers 
to match the \lya~forest redshift path of each quasar.
We then convolve the flux from the perfect skewers with the lifetime 
position-dependent HST-COS FUV line spread function for the G130M grating, 
interpolate it to match the wavelength vectors of the real data (with a velocity 
pixel scale of $\Delta v \approx 6$ km s$^{-1}$), 
and add Gaussian random noise to each pixel based on the error vectors of the real data.

Once we have generated both perfect skewers and forward-modeled skewers, we use an 
automated Voigt profile fitting code called VPfit {v10.2} \citep{Carswell14}\footnote{For 
more information, see \href{http://www.ast.cam.ac.uk/~rfc/vpfit.html}
{http://www.ast.cam.ac.uk/~rfc/vpfit.html}} to decompose each \lya~absorption spectrum 
into a collection of absorption lines. Each absorption line is described by three 
parameters: the absorption redshift, the Doppler width ($b$ in km s$^{-1}$), and the 
\HI~column density (\nhi~in cm$^{-2}$). To fit the skewers using VPfit, we use a fully 
automated Python wrapper that operates VPfit, as described in \citet{Hiss18, Hiss19} and 
also used for analysis in \citet{Teng22}. 

For comparison, in Fig.~\ref{fig.skewers} we show a chunk of \lya~forest spectrum 
generated from a line-of-sight passing through the 
same spatial coordinates in the Illustris and IllustrisTNG simulations. 
The choice of $\Gamma_{\rm HI}$ values used for this are
explained in section \ref{sec.gamma}. 
For visualization purposes, the figure shows only a small portion of the spectra
($\pm 750 $ km s$^{-1}$; $\sim \pm 10$ pMpc) at the center of the simulation box.
The top panel shows our forward-modeled skewers as well as the perfect skewers.
Blue ticks indicate the locations of  \lya~absorption lines automatically 
identified by VPfit when applied to the forward-modeled skewers, and numbers in 
the parenthesis indicate the $b$ (km s$^{-1}$) and $\log N_{\rm HI}$ (cm$^{-2}$)
obtained by VPfit.
In this example, only two \lya~absorption lines are 
identified and fitted by VPfit in the small portion of the skewers shown. 

The figure also shows temperature and density, and the corresponding optical 
depth weighted quantities along the skewers.
Optical depth weighted overdensities $\Delta^{\tau}_i$ and temperatures $T^{\tau}_{i}$ at
pixel $i$ are calculated as
\begin{equation}
\Delta^{\tau}_i = \frac{\sum_{j=1}^{n} \tau_{ij} \Delta_j}{\sum_{j=1}^{n} \tau_{ij} } 
\quad \&  
\quad T^{\tau}_i = \frac{\sum_{j=1}^{n} \tau_{ij} T_j}{\sum_{j=1}^{n} \tau_{ij} },
\end{equation}
where $\tau_{ij}$ is the optical depth contribution of real-space pixel $j$ 
to the total redshift-space optical 
depth at pixel $i$, and $n$ is the total number of pixels in each spectrum. 
The optical depth weighted density and temperature
are thus redshift space quantities that can be compared directly
to the redshift space optical depth. 

Since both simulations are run with the same initial conditions, the differences in 
the $T$,  $\Delta$, and line-of-sight velocity ($v_z$) along the skewer
are the result of the different feedback prescriptions. 
The temperature and $v_z$ panels for the region on the left side of the skewers 
show that the gas is hotter in 
Illustris as compared to IllustrisTNG, and the hotter regions of Illustris have steeper
velocity gradients as compared to IllustrisTNG. 
The explosive feedback is partly responsible for the extra velocity kinks seen 
in the Illustris velocity profile as compared to the smoother one in IllustrisTNG
(see $v_z$ panel of Fig.~\ref{fig.skewers}). 
The bottom panel of  Fig.~\ref{fig.skewers} 
shows the H~{\sc i} column density at each pixel (of $\sim 60$ kpc) in real space. 
When the bottom panel is compared to other panels, it illustrates that the high 
$N_{\rm HI}$ gas ($\gtrsim 10^{13}$) corresponding to high density ($\Delta > 3$ ) 
and lower temperature gas ($T \sim 10^4$ K; see the optical depth-weighted density 
and temperatures) imprints detectable 
absorption lines on the spectra. Note that the forward modeled spectra shown here are 
generated using a constant noise vector with a signal-to-noise ratio of 7 per pixel 
(of  $\Delta v \sim 6 $ km s$^{-1}$).

By running VPfit on these mock spectra we can compare the distribution of 
$b$ and $N_{\rm HI}$ in both simulations. 
However, the number of absorption lines and $N_{\rm HI}$ depend on the photoionization
rate $\Gamma_{\rm HI}$ from the ionizing UV background radiation. 
Therefore, when we want to study the effect of feedback on the IGM, 
$\Gamma_{\rm HI}$ becomes a nuisance parameter that we need to fix for both 
simulations. In the next section, we will discuss in more detail how to fix 
this nuisance parameter.

\section{Fixing the nuisance parameter $\Gamma_{\rm HI}$}
\label{sec.gamma}

The statistical properties of the IGM, such as the mean flux of
the \lya~forest, the flux power spectrum, $b - N_{\rm HI}$ distribution
or column density distribution, are 
important tools for understanding the IGM. These properties depend not only on the 
temperature-density distribution of the gas in the simulation 
but also on the $\Gamma_{\rm HI}$
sourced by the UV background. To disentangle the impact of the physical properties of the
simulated gas distribution from the impact of the UV background, it is necessary to 
fix the $\Gamma_{\rm HI}$. This is why it is 
common practice to consider $\Gamma_{\rm HI}$ as a nuisance parameter and fit for it 
when analyzing simulated \lya~forest data. 
To set the value of $\Gamma_{\rm 
HI}$, simulations must match one or more of the statistical properties of the observed 
IGM. For example, in high-redshift IGM studies, researchers often use the mean flux in 
the \lya~forest as a reference. 
In low-redshifts \citet{Viel17} used H~{\sc i}~column density distribution (CDD; defined in Eq.~\ref{eq.cdd}) to tune the $\Gamma_{\rm HI}$.
In this work, we have chosen to use the density of 
\lya~lines defined as the number of \lya~lines per unit redshift interval 
(${\rm dN/dz}$) as our reference statistic for determining the value of 
$\Gamma_{\rm HI}$. The range of column densities is chosen to be in between 
$12< \log \, N_{\rm HI} \,\,({\rm cm}^{-2}) < 14.5$ for calculating ${\rm dN/dz}$. 
Matching ${\rm dN/dz}$ is similar to the approach of matching the mean flux or 
effective \lya~optical depth that is commonly used 
at higher redshifts ($z>2$). We have chosen not to use the mean flux as a 
reference because at low redshifts ($z<0.5$), the normalized mean flux 
in the \lya~forest is very close to unity, and hence uncertainties 
in the placement of the continuum can have a large impact on its measurement.

To determine the value of $\Gamma_{\rm HI}$ in both simulations, we first obtained the 
${\rm dN/dz}$ value from observations. To do this, we fit Voigt profiles to 
the \lya~forest in 34 quasar spectra in the redshift range of $0.06<z<0.16$ from the 
dataset of \citet{Danforth14} having median S/N $\geq 10$ per resolution element.
We chose the region of \lya~forest in each spectrum to lie in the rest-wavelength range
1050-1180 \AA~ to avoid contamination by Ly$\beta$ lines and to remove
the quasar's 
line-of-sight proximity zone region.
Our VPfit code identifies 369 lines in the column density range 
$12< \log \, N_{\rm HI}\,\, ({\rm cm}^{-2})\, < 14.5$  over the total \lya~redshift 
path $\Delta z = 1.792$\footnote{The redshift path is obtained by summing the
total redshift covered by the \lya~forest removing the regions where intervening metals 
and the metals from Milky-Way ISM are present.} 
covered by the data in the aforementioned redshift range. 
This gives us the number density of \lya~lines ${\rm dN/dz} =205.9$. 
Using the simulations, we created a large number ($\approx 3500$) of forward modeled 
\lya~forest spectra for different values of $\Gamma_{\rm HI}$ (from $5\times 10^{-15}$ 
to $2\times 10^{-13}$ s$^{-1}$) following the procedure described in the previous section. 
We then fit these skewers 
with the same automated VPfit code, storing the values of \nhi~and $b$ for each 
\lya~line and determining the ${\rm dN/dz}$ for each $\Gamma_{\rm HI}$. 
%
%
\begin{figure}
\includegraphics[width=0.48\textwidth,keepaspectratio]{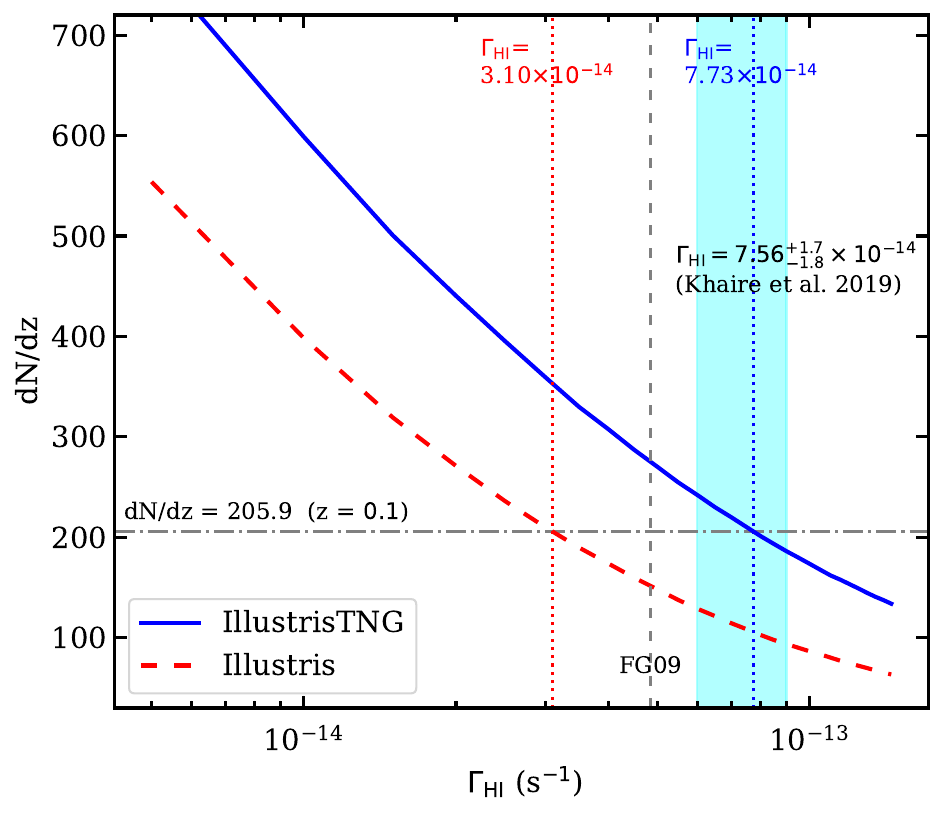}
\caption{ The line density ${\rm dN/dz}$ for \lya~absorbers in the range 
$10 ^{12} < N_{\rm HI} < 10^{14.5} \, { \rm cm^{-2} }$ is plotted for 
IllustrisTNG (blue curve) and Illustris (red dashed curve) simulations at $z=0.1$. 
The grey dot-dashed line shows ${\rm dN/dz}= 205.9$, 
which was obtained by fitting Voigt 
profiles to the \lya~forest observed in HST COS data from \citet{Danforth14} 
in the redshift range $0.06<z<0.16$. The vertical dotted lines indicate the values of 
$\Gamma_{\rm HI}$ ($3.1 \times 10^{-14}$ s$^{-1}$ for Illustris and $7.73 \times 
10^{-14}$ s$^{-1}$ for IllustrisTNG) at which the simulations match the ${\rm dN/dz}$ 
of the data. The cyan-shaded region shows the measurements of $\Gamma_{\rm HI}$ and 
its 1$\sigma$ uncertainty obtained by \citet{Khaire19} using Nyx  
simulations without any feedback. The gray dashed vertical line shows the $\Gamma_{\rm HI}$ 
at $z=0.1$ from \citet{FG09} used by both Illustris and IllustrisTNG. 
This figure demonstrates that Illustris, which has 
explosive feedback and a small fraction of gas mass in the diffuse \lya~phase, 
requires a lower value of $\Gamma_{\rm HI}$ to match the observed ${\rm dN/dz}$.
 }
\label{fig.gama}
\end{figure}

In Fig. \ref{fig.gama}, we present the values of ${\rm dN/dz}$ obtained for various 
$\Gamma_{\rm HI}$ values in both IllustrisTNG and Illustris simulations. The 
horizontal line, indicating our measurement of ${\rm dN/dz} =205.9$ from the 
data, serves as a reference for comparison. Our results show that 
IllustrisTNG requires $\Gamma_{\rm HI} = 7.73 \times 10^{-14}$ s$^{-1}$ to match this 
measurement, while Illustris requires $\Gamma_{\rm HI} = 3.1 \times 10^{-14}$ s$^{-1}$ 
(as indicated by the vertical dotted lines in Fig.~\ref{fig.gama}).
We use these $\Gamma_{\rm HI}$ values for 
comparing the results of Illustris and IllustrisTNG
throughout this paper. Note that the example skewer shown in
Fig.~\ref{fig.skewers} has been calculated for these 
values of $\Gamma_{\rm HI}$ only.

Here we have used  ${\rm dN/dz}$ in $ 12 < {\rm log}N_{\rm HI} \,\, ({\rm cm}^{-2})<14.5$ range 
for tuning the $\Gamma_{\rm HI}$. However, if we choose a small interval $ 13 < {\rm log}N_{\rm HI} \,\, ({\rm cm}^{-2}) <14.5$ which is unaffected by the incompleteness of line identifications in \citet{Danforth14}, we get $\Gamma_{\rm HI}$ that is
15\% higher than those obtained here. 
Instead of using ${\rm dN/dz}$  for tuning 
$\Gamma_{\rm HI}$ one can use either mean flux or H~{\sc i}~CDD. These are all valid approaches
\citep[see e.g][]{Viel17, Bolton_Gaikwad22} 
however one needs to be cautious while using mean 
flux at low-$z$ because it is very close to 1 where the noise as well as the continuum placement uncertainty can severely affect its measurements. Instead of 
$\rm dN/dz$ if we use the mean flux for tuning $\Gamma_{\rm HI}$, calculated as the mean of the continuum normalized flux in the \lya~forest region excluding the metal lines in the same redshift bin $0.6<z<0.16$, we obtain $\Gamma_{\rm HI}$ values that are within 5\% of those obtained using $\rm dN/dz$. In Section~\ref{sec.cdd},
we demonstrate that our $\Gamma_{\rm HI}$ values obtained through ${\rm dN/dz}$ matching yield CDD consistent with observations within the same $12 < {\rm log}N_{\rm HI} \,\, ({\rm cm}^{-2})\, < 14.5$ range used ${\rm dN/dz}$
(see Fig.~\ref{fig.cdd}). 
Alternatively, if we use the CDD within the $13 < {\rm log}N_{\rm HI} \,\, ({\rm cm}^{-2})\, < 14.5$ range, we find that the resulting values of  $\Gamma_{\rm HI}$ for both simulations are within 10\% of those obtained using ${\rm dN/dz}$ for the same $N_{\rm HI}$ range. Therefore we conclude that, for the purpose of calibrating $\Gamma_{\rm HI}$, the use of ${\rm dN/dz}$ is equivalent to using mean flux or CDD.

Given that the mass fraction in the diffuse
\lya~phase is quite different in the Illustris and IllustrisTNG simulations,
we expect the $\Gamma_{\rm HI}$ values to be different in order to match the 
${\rm dN/dz}$. This is a result of the fact that most of the summary statistics of the \lya~forest  (e.g. mean flux, CDD or $\rm dN/dz$) are 
related to the mean effective \lya~optical depth $\tau$ 
which can be approximated as,
\begin{equation} \label{eq.fgpa}
    \tau \propto \Gamma_{\rm HI}^{-1} \, T^{-0.7}_{0}({\rm f_{Ly\alpha}}\,\Omega_b \,h^{2})^2 \,\Omega_m^{-0.5},
\end{equation}
following the fluctuating Gunn-Peterson approximation \citep{Weinberg97} where ${\rm f_{Ly\alpha}}$
is the fraction of mass in the diffuse \lya~phase (see Fig.~\ref{fig.T-Delta}). 
As expected from Eq.~\ref{eq.fgpa}, the required $\Gamma_{\rm HI}$ for Illustris 
is smaller than IllustrisTNG because of the lower ${\rm f_{Ly\alpha}}$. 
The expected ratio of $\Gamma_{\rm HI}$ values of IllustrisTNG relative to Illustris 
(using Eq.~\ref{eq.fgpa})
is a factor of $2.52$ when we demand that the mean optical depths of both 
simulations should be the same and
use the ${\rm f_{Ly\alpha}}$ and cosmological parameters from table~\ref{tab.cosmo} and
\ref{tab.gas_phases}. Note that most of the difference arises from the 
${\rm f_{Ly\alpha}^2}$ factor. The ratio of $2.49$ that we obtained by  
by ${\rm dN/dz}$ matching is in excellent agreement with the expectation from this scaling argument based on the
fluctuating Gunn-Peterson approximation. 

Note that Illustris and IllustrisTNG used the \citet{FG09} UV background
which gives $\Gamma_{\rm HI} = 4.85 \times 10^{-14}$ s$^{-1}$ at $z=0.1$ as indicated by gray dashed vertical line in Fig.~\ref{fig.gama}.
Assuming this value of $\Gamma_{\rm HI}$, Illustris under-predicts
the measured ${\rm dN/dz}$ by $26 \%$ and IllustrisTNG over-predicts it by $34\%$.

In \citet{Khaire19}, we measured $\Gamma_{\rm HI}$ to be 
$7.56 ^{^{+1.7}_{-1.8}} \times 10^{-14}$ s$^{-1}$ at $z=0.1$ (as shown by the cyan shade 
in Fig.\ref{fig.gama}). For this measurement, we compared a measurement
of the \lya~forest flux power spectrum to an ensemble of Nyx 
simulations \citep{Almgren13}, which do not model galaxy formation or feedback. 
This value of $\Gamma_{\rm HI}$ is consistent with the value 
needed to match the ${\rm dN/dz}$ measurement for IllustrisTNG, but  $2.5 \, \sigma$ 
higher than the value needed for Illustris. This consistency between the Nyx 
and IllustrisTNG $\Gamma_{\rm HI}$  is not surprising, as the diffuse \lya~fraction in 
the Nyx simulation (39.3\%) is only 1\% higher than in  IllustrisTNG \citep[see][]{Hu23_whim}.

By utilizing the simple statistical measure 
${\rm dN/dz}$ to fix the nuisance parameter $\Gamma_{\rm HI}$, 
we have demonstrated that the ${\rm dN/dz}$ depends on $f_{\rm Ly\alpha}^2$/$\Gamma_{\rm HI}$, 
akin to the mean \lya~effective optical depth. This degeneracy between $f_{\rm Ly\alpha}$ and 
$\Gamma_{\rm HI}$ is critical to comprehend the relationship between feedback and the inferred 
$\Gamma_{\rm HI}$. If feedback removes gas from the diffuse \lya~phase by heating it to a higher 
temperature, thereby reducing $f_{\rm Ly\alpha}$, then the inferred $\Gamma_{\rm HI}$ needed to match 
the ${\rm dN/dz}$ will also decrease. This scenario is observed in simulations with extreme feedback, 
such as Illustris, which has a $f_{\rm Ly\alpha}$ of just $23$\% 
(see Table~\ref{tab.gas_phases}), compared to the $38$\% 
of IllustrisTNG. Therefore, we require different values of $\Gamma_{\rm HI}$ to compare the 
IGM from both simulations.

After adjusting the nuisance parameter $\Gamma_{\rm HI}$, we proceed to examine the impact of 
feedback on the \lya~forest in both simulations. Our findings will be presented and discussed 
in the following section.

\section{Quantifying the Impact of Feedback on IGM Statistics}\label{sec.igm}

We investigate the impact of feedback on the IGM by comparing a variety of statistics, 
including the $b$-distribution, the \nhi-distribution, the joint $b-$\nhi~distribution, 
the column density distribution function, and the flux power spectrum. 
The results of these are described below.
\begin{figure*}
\includegraphics[width=0.98\textwidth,keepaspectratio]{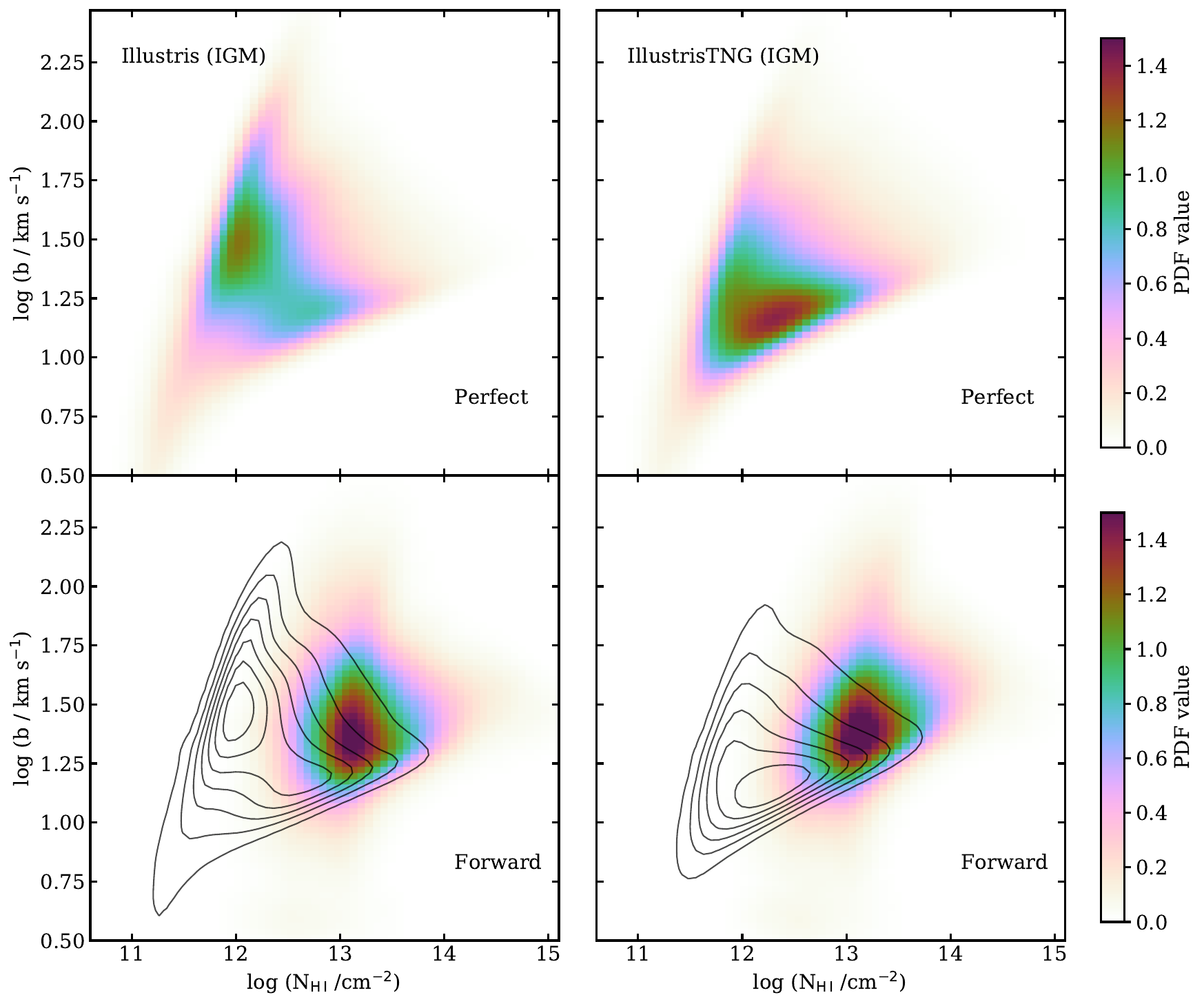}
\caption{The $b$-\nhi~distributions for Illustris (left panel) and
IllustrisTNG (right panel) are shown for perfect IGM skewers (top 
panels) and forward-modeled \lya~spectra (bottom panels). For better 
visual comparison, the contours of the perfect $b$-\nhi~distributions are 
also included in the bottom panels. The effect of feedback can be seen in 
the different shapes of the perfect $b$-\nhi~distributions (top-panels). 
The hotter gas in Illustris results in more gas at high $b$ low \nhi~values, 
while the slightly cooler gas in IllustrisTNG results in lower $b$ 
high \nhi~values. However, the $b$-\nhi~distributions for forward-modeled 
\lya~spectra are similar for both simulations. The difference seen in the 
perfect models is not present in the forward modeled $b$-\nhi. 
This is because the forward-modeled spectra are not sensitive to these 
lower \nhi~values where perfect models show the difference.
}
\label{fig.bN_igm_perfect}
\end{figure*}
\begin{figure*}
\includegraphics[width=0.98\textwidth,keepaspectratio]{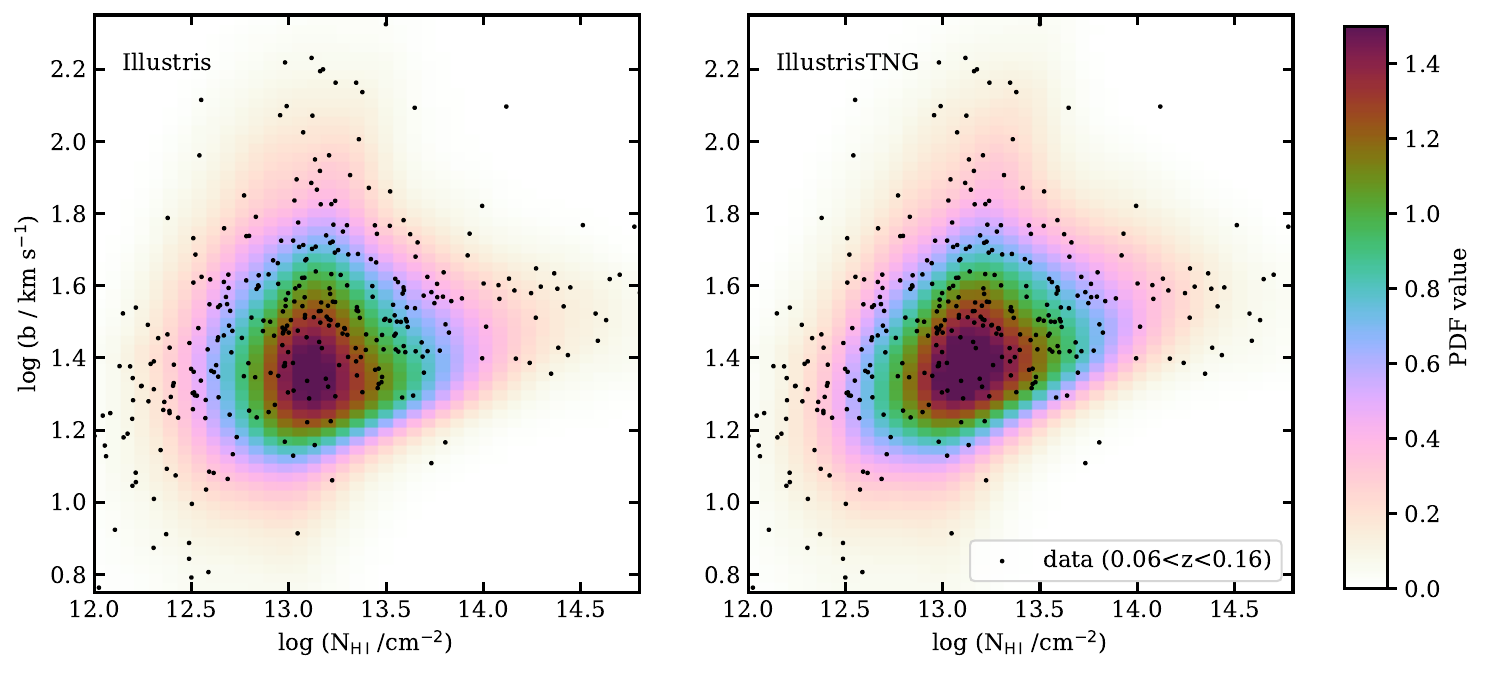}
\caption{The $b$-\nhi~distribution from forward modelled \lya~forest 
(as shown in the bottom panel of Fig.~\ref{fig.bN_igm_forward})
for Illustris (left) and IllustrisTNG (right)
along with the $b$-\nhi~values obtained by Voigt profile fitting the 
\lya~forest observed in HST COS data from \citet{Danforth14} in a redshift bin $0.06<z<0.16$. Both simulations under-predict the $b$ values (see also Fig.~\ref{fig.bN_igm_separate} for separate $b$ and \nhi~distribution).  
}
\label{fig.bN_igm_forward}
\end{figure*}
\begin{figure*}
\includegraphics[width=0.98\textwidth,keepaspectratio]{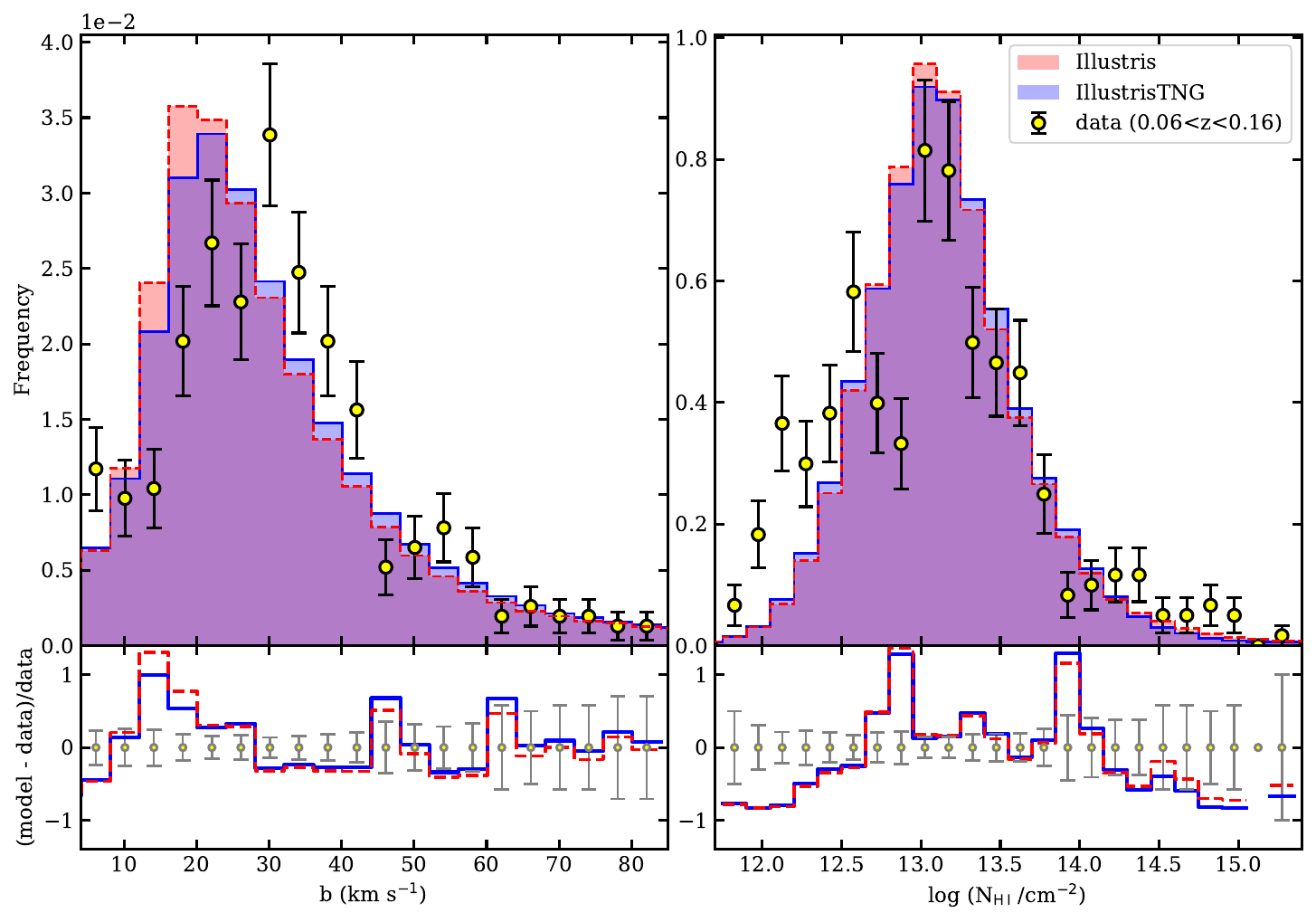}
\caption{Normalized distribution of $b$ (left-hand panel) and \nhi~(right hand panel) for 
forward modelled \lya~forest skewers from Illustris 
(reds-dash histogram) and IllustrisTNG (blue histogram).
Forward models from both simulations provide similar $b$ and \nhi~distributions. Data points
show the normalized $b$ and \nhi~distribution obtained by fitting the 
\lya~forest in \citet{Danforth14} data in a redshift bin $0.06<z<0.16$. The bottom panel 
shows the fractional difference between the
histograms from simulations and the data, while error bars show the 
the fractional difference in the uncertainty of the data. 
Both simulations fail to reproduce the $b$ distribution at $b\lesssim 45$ km s$^{-1}$ and 
\nhi~distribution at low (\nhi $\, <10^{12.5}$ cm$^{-2}$) 
and high \nhi($\,>10^{14.5}$ cm$^{-2}$).}
\label{fig.bN_igm_separate}
\end{figure*}
\subsection{The $b-$\nhi~distribution}
To obtain the $b-$\nhi~distribution from simulations, we need to Voigt profile fit the 
mock \lya~forest spectra created from those simulations. However, the automated Voigt 
profile fitting code, VPfit, does not work well on the perfect, noiseless skewers. 
This is mainly because VPfit tries to identify absorption lines in regions that 
are bounded by flux equal to the continuum. However, for most of the redshift path, the continuum normalized flux is 
slightly below unity due to the highly ionized IGM at low 
redshifts.\footnote{The differences in the flux and continuum are of the order of 
$10^{-3}$ - $10^{-5}$.} As a result, VPfit fails to identify and fit the lines in 
these regions. To mitigate this issue, as done previously in \citet{Hiss18} and \citet{Teng22}, 
we add Gaussian random noise to the flux of 
the perfect skewers so that the S/N per pixel ($\Delta v = 4.1$ km s$^{-1}$) is equal to 100 
(the equivalent of $\sim 200$ per resolution element for HST COS spectrum). 
This allows VPfit to accurately identify \lya~lines and fit to obtain $b$ and \nhi~for 
each line. 

We ran VPfit on $10^5$ perfect skewers (after adding a tiny amount of noise as mentioned above)
and obtain the values of  $b$ and \nhi.
Using these values, we compute a Kernel 
Density Estimate (KDE) of the joint
$b-$\nhi~distribution using a Gaussian kernel, which is shown in the top panels of 
Fig.~\ref{fig.bN_igm_perfect} for Illustris (left panel) 
and IllustrisTNG (right panel). 
Clear differences can be seen in the shape of the 
$b$-\nhi~distribution. 
In Illustris, most of the lines fall in the low column density (\nhi$< 10^{12.3}$ 
cm$^{-2}$) region with higher $b$ values ($b>$ 18 km s$^{-1}$). 
In contrast, in IllustrisTNG, the gas is distributed more uniformly across the 
$b-$\nhi~plane, with more gas at high column densities (\nhi$ > 10^{11.8}$ cm$^{-2}$) 
and low $b$ values ($b<$ 20 km s$^{-1}$). 
The differences in the $b$ values indicate that the 
\lya~absorbing gas is hotter in Illustris compared to IllustrisTNG. 
However, differences in the column densities, 
where IllustrisTNG shows more gas at slightly 
higher column densities than Illustris, are not significant, as they are expected to 
be similar due to the way we fixed the values of $\Gamma_{\rm HI}$.
It is interesting to see that, even after fixing the nuisance parameter $\Gamma_{\rm HI}$
to give the same $\rm dN/dz$, the effect of feedback is clearly seen in the  
shape of the $b-$\nhi~distribution 
of the perfect skewers. Therefore, ideally one can use 
very high S/N spectra ($\sim 100$ per pixel) and resolution 
($\sim 4.1$ km s$^{-1}$) for constraining the effect of feedback on the IGM,
however unfortunately 
only a small fraction of data would exist close to this S/N and resolution (e.g. using 
the Space Telescope Imaging Spectrograph). 

To determine whether the differences in the $b$-\nhi~distribution observed in the 
perfect skewers can be detected in real data, we conducted the same analysis on our 
forward-modeled mock spectra. We created these models using the noise properties of 
\citet{Danforth14} for the same $10^5$ perfect skewers. On average, we needed to stitch 
around three skewers together for each forward-modeled spectrum, resulting in a total of approximately 
35,000 forward-modeled spectra. In contrast to the noiseless perfect skewers, VPfit works well 
on these forward models because of the noise. The KDEs of the joint $b-$\nhi~distribution 
from these forward models are shown in the bottom panels of 
Fig.~\ref{fig.bN_igm_perfect} for Illustris (left panel) and IllustrisTNG 
(right panel). The figure shows that the differences between the two 
simulations that were present in the $b-$\nhi~distribution of the perfect skewers 
(shown in the lower panels of Fig.~\ref{fig.bN_igm_perfect} by contours) are completely washed out in the 
forward models. The joint $b-$\nhi~distribution looks similar for both Illustris and 
IllustrisTNG, which is mainly because only the low column density systems 
($10^{11.5} <\, $\nhi$ \, < 10^{13.2}$ cm$^{-2}$) showed differences in the case of the 
perfect skewers. These low-column density lines are not 
detected in the forward modeled HST/COS spectra given the lower S/N and moderate resolution ($R\sim 15000-20000$). The normalization of the joint 
$b-$\nhi~distribution (as indicated by color-bar) is also the same for 
both simulations, as expected due to our choice of $\Gamma_{\rm HI}$ to match ${\rm 
dN/dz}$. This suggests that even the best available dataset of low-$z$ 
\lya~forest, as provided in \citet{Danforth14}, can not probe the effects of 
feedback on the IGM using the $b-$\nhi~distribution. 

To further explore this, in Fig.~\ref{fig.bN_igm_forward}, we compare the 
$b-$\nhi~distribution from Illustris and IllustrisTNG with the observations in the 
$0.06 < z < 0.16$ bin that we fit using the same automated VPfit code. 
The data appears to agree reasonably well with both simulations in terms of \nhi, but 
observations show an excess of lines with higher $b$ values. 
To further investigate this discrepancy, we plot the normalized marginalized histograms of $b$ and \nhi~in Fig.~\ref{fig.bN_igm_separate}. 
It can be seen that both simulations fail to match the $b$ distribution, 
with peaks at $b \approx 20$  km s$^{-1}$ for Illustris and IllustrisTNG, 
while the peak in the data appears at $b\approx 30$ km s$^{-1}$.
Previous studies  \citep[][]{Viel17,Gaikwad17b, Nasir17} 
have also noted this  discrepancy in $b$ values, which has been argued to indicate a hotter IGM or
some effect of feedback or 
other possible sources of additional heating such as from dark photon dark matter \citep[see][]{Bolton22}. 
However, even in the case of extreme feedback 
in Illustris, the discrepancy remains. 
It seems that Illustris even gives a slightly smaller peak in $b$ than 
IllustrisTNG, although the difference is not 
significant.\footnote{
We also note that the $b$-distribution in both simulations is almost identical even if we use the same $\Gamma_{\rm HI}$ in both. However, see \citet{Mallik23b} for the effect of feedback and $\Gamma_{\rm HI}$ on $b$-distribution of aligned absorbers.} 
However, it does not definitively mean that feedback can not 
affect the $b$ distribution and it might be that feedback models implemented in both Illustris and IllustrisTNG simulations are lacking some ingredients to reproduce the observed $b$ distribution.
On the other hand, by construction given the ${\rm dN/dz}$ matching, 
both simulations give almost identical 
\nhi~distributions in the  $12 < {\rm log N_{HI}} \,\, ({\rm cm}^{-2})<14.5$ range. 
There is a mild disagreement around two bins centered at 
log \nhi~$\approx 12.7$ and $13.3$  where both simulations overproduce the number
of \lya~lines. 
At higher column densities (log \nhi~$>15$) simulations under-produce \lya~lines.

In summary, we find that Illustris and IllustrisTNG exhibit 
different  $b-$\nhi~distributions, however, the existing data 
does not have a sufficiently high S/N ratio and resolution to discern this 
difference. Both simulations are also unable to reproduce the discrepancy between the 
observed and simulated $ b$ distributions at low redshifts. This suggests the need for 
further investigation into the issue of the $ b$ distribution 
\citep[see][]{Bolton_Gaikwad22, Bolton22}.

\subsection{The column density distribution}\label{sec.cdd}
\begin{figure}
\includegraphics[width=0.48\textwidth,keepaspectratio]{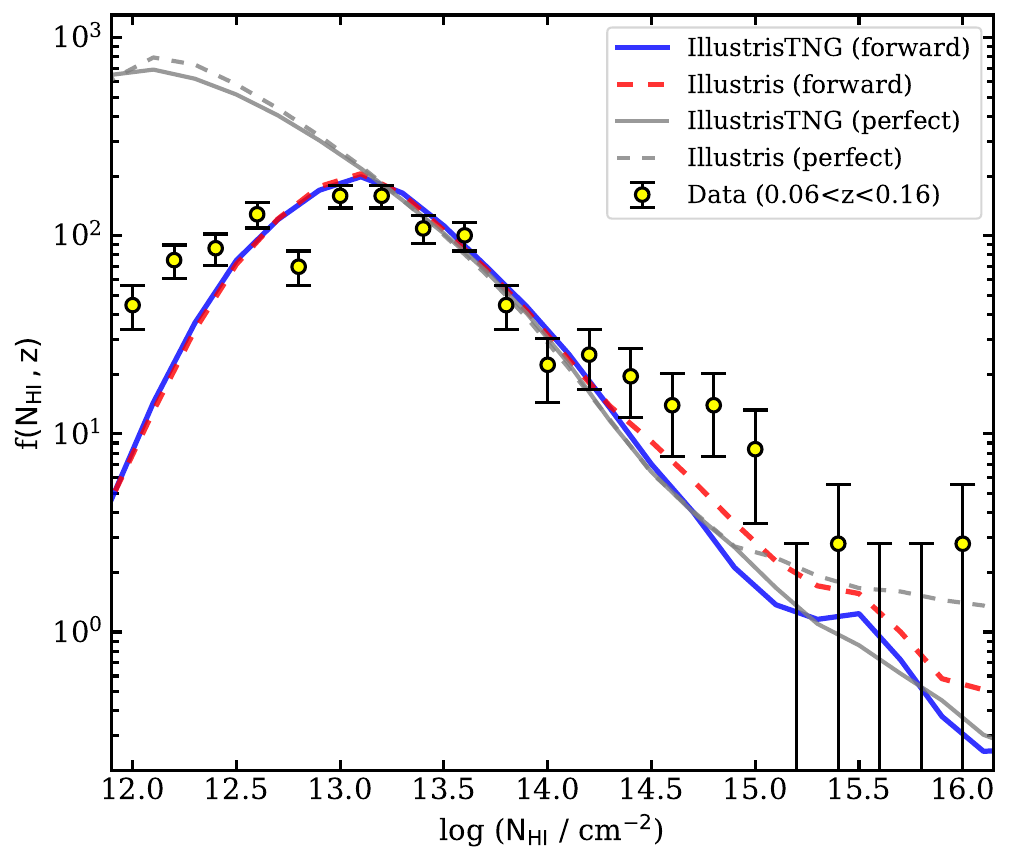}
\caption{ 
The figure shows the $z=0.1$ H{\sc ~i}~column density distribution (CDD) for 
Illustris (dashed curves) and IllustrisTNG (solid curves) simulations
when we use perfect and forward-modeled skewers. We also show the CDD 
obtained by fitting the \citet{Danforth14} data at $0.06<z<0.16$ bin using 
our VPfit code.  The simulations show good agreement with each other at
log \nhi$\lesssim \, 14.5$ but diverge at higher column densities. 
In forward models, both simulations are consistent with CDD measurement 
at $12.3 < \log \, N_{\rm HI}< 14.5$ but under-predict it at higher column 
densities, with IllustrisTNG showing a greater 
under-prediction than Illustris.}
\label{fig.cdd}
\end{figure}

We calculate the H{\sc ~i}~CDD defined as
\begin{equation}\label{eq.cdd}
    f(N_{\rm HI}, z) = \frac{d^2N}{d \log N_{\rm HI} \,dz} ,
\end{equation}
where $N$ is the number of \lya~absorption lines with H{\sc ~i} column density in the
range $\log N_{\rm HI}$ to $\log N_{\rm HI}$ + $\Delta \log N_{\rm HI}$ detected
within the observed redshift pathlength $dz$. We calculate the CDD for perfect skewers as
well as forward-modeled skewers for both simulations. The results are shown in 
Fig.~\ref{fig.cdd}. 

Fig.~\ref{fig.cdd} shows that both simulations give identical CDDs for 
$\log N_{\rm HI} \lesssim 14.5$ for perfect 
skewers (gray curves) as well as forward-modeled skewers (blue and red curves). 
This is not surprising given our method of fixing the nuisance parameter
$\Gamma_{\rm HI}$. The CDDs of Illustris and IllustrisTNG
begin to deviate from each other at 
$\log N_{\rm HI} \,\, ({\rm cm}^{-2}) \gtrsim 15$ and $\log N_{\rm HI} \,\, ({\rm cm}^{-2}) \gtrsim 14.3$  
where Illustris gives higher CDD than IllustrisTNG for perfect 
skewers and forward-modeled skewers, respectively. 
This small deviation may be related to the feedback prescriptions
as well as how VPfit fits noisy, saturated lines. Given that Illustris generally 
has more shock-heated gas than IllustrisTNG, the lines are wider in Illustris 
and therefore saturate at slightly higher column densities than for Illustris. 
This leads to more lines and hence a higher CDD at higher column densities for Illustris 
compared to IllustrisTNG. 
According to this argument if the data has an overall higher S/N  then the 
$N_{\rm HI}$ above which the CDD deviates in both simulations would be higher. 
This is what is seen in the case of the perfect skewers that have S/N of 100 per
pixel (of size $4.1$ km s$^{-1}$, because of added noise to the spectra) as compared to forward models. 
There is also a small
difference in the CDD of Illustris and IllustrisTNG 
at lower column densities (\nhi~$<10^{12.5}$ cm$^{-2}$) for perfect skewers. This difference is too small to 
be detected in a realistic dataset and becomes even smaller for the forward model spectra. 

Fig.~\ref{fig.cdd} depicts a notable decrease in the CDD at 
$\log \, N_{\rm HI} \lesssim 13$ for forward-modeled skewers as opposed to perfect skewers. 
This reduction is attributed to the undetected lower column density \lya~lines that are obscured 
by the noise. Consequently, noisy forward-modeled skewers yield lower CDD values in comparison to 
perfect skewers (with S/N = 100 per pixel of size $4.1$ km s$^{-1}$). The minimum column density at which the downturn occurs depends 
on the overall S/N ratio of the data. In typical CDD measurements, such as those conducted 
by \citet{Danforth14}, this downturn is often corrected by a completeness correction that accounts 
for the number of lines that can be missed in the noisy data. As we forward model the spectra using 
the real data, we can conduct a direct comparison of the CDD with observations, rendering such 
completeness correction unnecessary. 

In Fig.~\ref{fig.cdd} we also show the CDD  that we obtain by fitting the 
\citet{Danforth14} \lya~forest in the $0.06<z<0.16$ bin. The CDD is calculated
in bins of $\Delta \log \, N_{\rm HI} = 0.2$, and errors on CDD only account for
the Poisson counting errors (i.e $\sqrt{N}$) on the number of lines in each bin. 
If there are no
lines in the bin then the upper limit on the confidence interval in counting lines
is taken as 1.
Since, as mentioned earlier, we will be comparing 
it with our forward models, we do not need to perform a completeness correction. 
As expected, our forward-modeled CDD in both simulations matches well with 
observations in $12 <\log N_{\rm HI} \,\, ({\rm cm}^{-2})< 14.5$ range. At higher column densities
both simulations under-predict the CDD. It is worth noting that Illustris provides a better fit to the data than 
IllustrisTNG at higher column densities $14 <\log N_{\rm HI} \,\, ({\rm cm}^{-2}) <15$. 
However, this tiny improvement in fitting the CDD can not be
treated as definitive evidence that the feedback model in Illustris is 
closer to reality than IllustrisTNG.

Overall, we conclude that the CDD from Illustris and IllustrisTNG simulations 
does not show any clear evidence of the effect of feedback on the IGM at 
 $\log N_{\rm HI} \,\, ({\rm cm}^{-2}) \lesssim 15$. 
 There is a small variation in the CDD at $\log N_{\rm HI} \,\, ({\rm cm}^{-2}) \gtrsim 15$,
 which may suggest that the explosive feedback in Illustris is responsible for the 
 higher CDD. However, the difference between the simulations and data is not 
 significant enough to favor any particular feedback model.

It is important to note that a large difference in the CDD, especially in its 
normalization, can be observed if $\Gamma_{\rm HI}$ is not calibrated. 
This could be 
misinterpreted as a potential observable to differentiate between feedback models
\citep[see for e.g][ and discussion in section~\ref{sec.disc}]{Burkhart22}.
But because we cannot predict $\Gamma_{\rm HI}$ from first principles, it must be 
treated as a nuisance parameter that can be fixed by matching the ${\rm dN/dz}$.
 Any comparison with observational data would be inaccurate if the degeneracy stemming from 
$\Gamma_{\rm HI}$ is not resolved. 
The only relevant aspect of the CDD for determining the best feedback models may be 
its shape, which is mostly independent of the value of $\Gamma_{\rm HI}$. 

\subsection{The \lya~Flux Power Spectrum}\label{sec.ps}

The \lya~flux power spectrum is a widely-used statistical measure for studying the IGM.
It has been employed to determine the temperature of the 
IGM \citep[e.g][]{Theuns00, Boera19, Walther19, Gaikwad21} and 
the UV background \citep[e.g][]{Gaikwad17a, Khaire19}, as 
well as to constrain cosmological parameters such as neutrino masses 
\citep[e.g][]{McDonald06, Yeche17Neutrino}
and to probe alternative dark matter models such as warm or fuzzy dark matter 
\citep[][]{Viel13, Irsic17t} at high redshifts. 
In order to investigate whether the \lya~flux power spectrum can be used to study the 
effect of feedback on the IGM, we compute it using $10^5$ perfect skewers from 
Illustris and IllustrisTNG. We first calculate the flux contrast, $\delta_F = (F - 
\bar{F})/\bar{F}$, and then compute the Fourier transform, $\hat{f}(k)$, of 
$\delta_F$. The power spectrum is then
defined as $P(k) \propto \, \mid \hat f(k) \mid ^2 $ with standard normalization
$\sigma^2_{\delta_F}=\int_{-\infty}^{\infty}{dk \, P(k)/2\pi}$.

\begin{figure}
\includegraphics[width=0.49\textwidth,keepaspectratio]{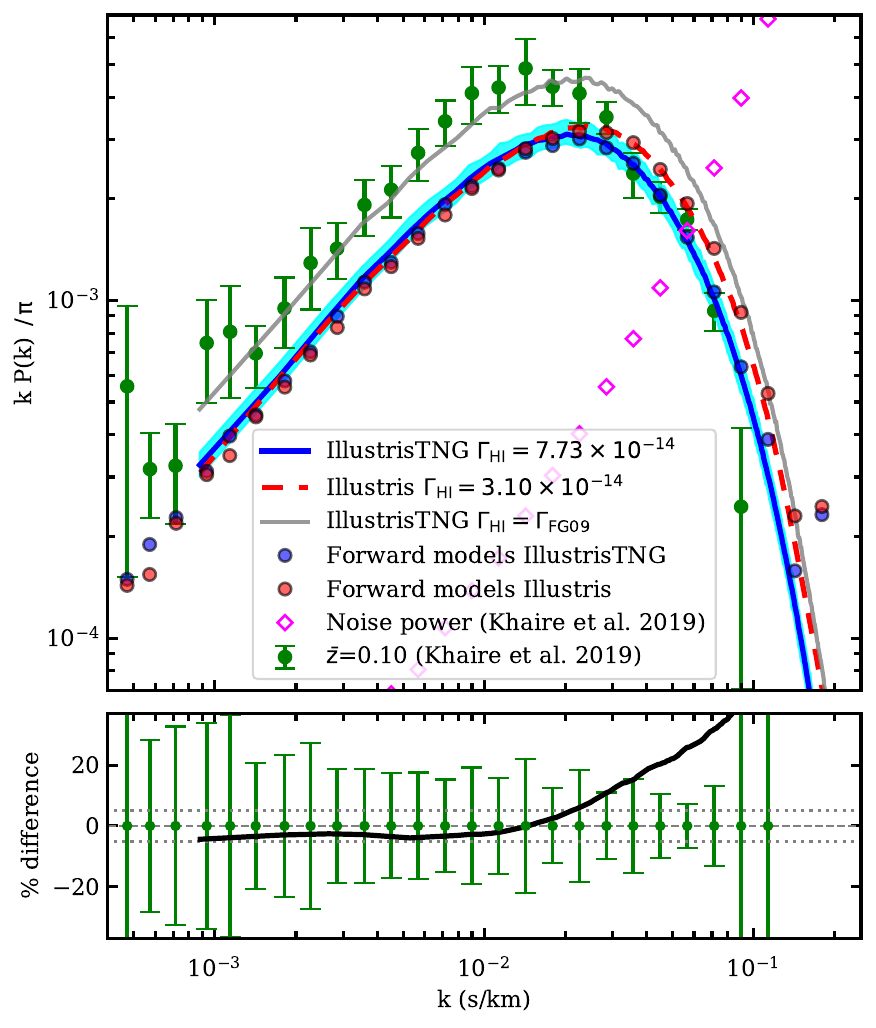}
\caption{Figure shows
line of sight \lya~flux power spectrum ($kP(k)/\pi$) obtained using perfect skewers from 
IllustrisTNG (blue curve) 
and Illustris (red-dash) at $z=0.1$ for the $\Gamma_{\rm HI}$ values obtained by matching 
${\rm dN/dz}$ along with the measurements 
(green data points) and the noise power (magenta diamonds) from \citet{Khaire19}. 
The red and blue circles show the power spectrum calculated from forward modeled
HST-COS spectra for Illustris and IllustrisTNG, respectively, following 
the same procedure employed for measurements in \citet{Khaire19}.  
The bottom panel shows the relative percentage difference (black curve)
in the power spectrum of Illustris (red-dash) with respect to IllustrisTNG (blue solid)
and the error bars show percentage errors in the measurements where dotted horizontal
lines are for $\pm 5$\%.
The discrepancy between simulations and measurements is discussed in 
Section~\ref{sec.ps}.
For comparison, we show the power spectrum from IllustrisTNG 
at the $\Gamma_{\rm HI}$ values of \citet{FG09}, 
i.e $\Gamma_{\rm FG09} = 4.85 \times 10^{-14}$ s$^{-1}$
(grey solid curve). The cyan shade corresponds to the range of power-spectrum 
from IllustrisTNG perfect skewers, when we take into account the 
Poisson error on $\rm dN/dz$ to obtain the $\Gamma_{\rm HI}$. 
}
\label{fig.ps}
\end{figure}

In Fig.~\ref{fig.ps}, we present the power spectrum ($kP(k)/\pi$) at $z=0.1$ 
calculated using the noiseless perfect skewers
for both 
Illustris and IllustrisTNG in the top panel, and the percentage difference between them 
in the bottom panel. The minimum wavenumber $k$ 
probed by the power spectrum is determined by the 
size of the skewers, which in this case is the size of the simulation box. At large 
scales  ($k< 0.02$ s km$^{-1}$), the power spectra from both simulations show good 
agreement. However, they begin to diverge at smaller scales ($k> 0.02$ s km$^{-1}$) where 
Illustris exhibits higher power than IllustrisTNG. The difference between the two 
simulations is approximately 5\% for $k < 0.02$ s km$^{-1}$ and increases monotonically 
at higher $k$ values. This difference is likely due to the impact of feedback on the IGM 
in the simulations. We compare the measurement uncertainties from the power spectrum 
measurements at $z=0.1$ by \citet{Khaire19} with this difference in the bottom panel of 
Fig.~\ref{fig.ps}. It shows that only at $k > 0.04$ s km$^{-1}$ is the difference in the 
simulations more than the measurement uncertainties. 
Therefore, the difference in the power spectrum at 
small scales (i.e. $k > 0.02$ s km$^{-1}$) may indicate the effect of feedback.
However, note that at these $k$ values, the 
noise power from the data begins to dominate the power spectrum measurements (see magenta 
diamonds in the top panel). 

Power spectrum measurements at $z=0.1$ plotted in Fig.~\ref{fig.ps} are from
our previous work \citet{Khaire19} where we used the same  dataset and redshift bin
for \citet{Danforth14} data to obtain the power spectrum. Contrary to our expectation, 
the power spectra from both simulations do not match the measurements at the
large scales ($k<0.02$ s km$^{-1}$). The simulated power spectra on average
under-predict the measurement at these scales by a factor of $\sim 1.5$. 
This huge discrepancy is surprising and unexpected. 
\begin{figure*}
\begin{subfigure}{0.5\textwidth}
  \centering
  \includegraphics[width=.99\linewidth]{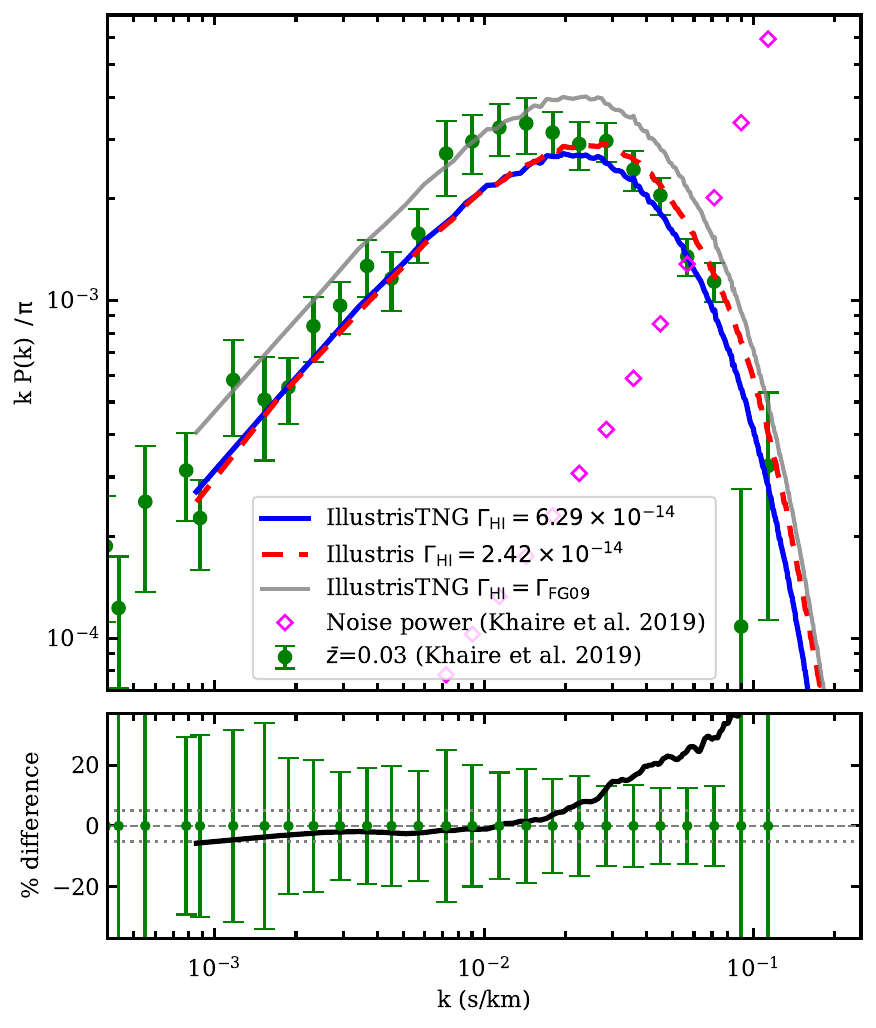}
  \label{fig:image1}
\end{subfigure}%
\begin{subfigure}{0.5\textwidth}
  \centering
  \includegraphics[width=.99\linewidth]{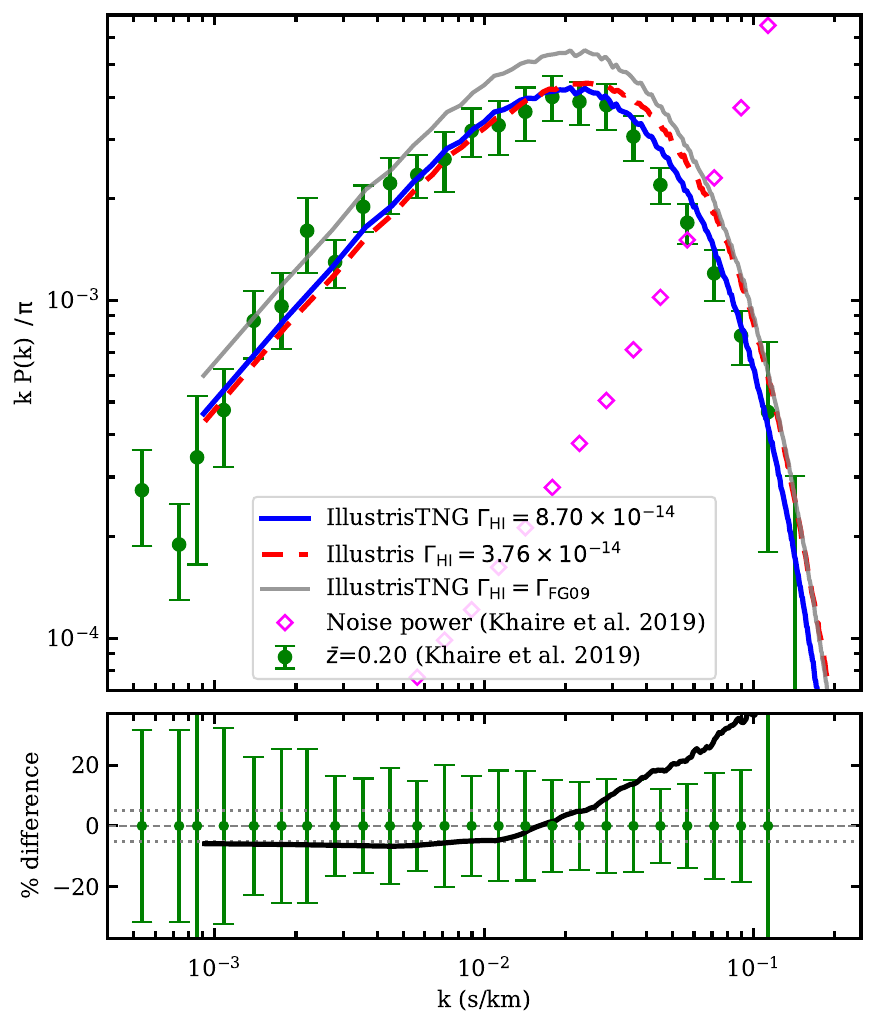}
  \label{fig:image2}
\end{subfigure}
\caption{Figure shows
line of sight \lya~flux power spectrum obtained using perfect skewers from 
IllustrisTNG (blue curve) 
and Illustris (red-dash) at $z=0.03$ (left-hand panel) and $z=0.2$ 
(right-hand panel) for the $\Gamma_{\rm HI}$
values (as indicated in the legends) obtained by ${\rm dN/dz}$ matching.
At these redshifts, the simulated power spectrum is consistent with measurements. 
The curves and the description of the figure are the same as Fig.~\ref{fig.ps} and 
$\Gamma_{\rm FG09}$ is $3.97 \times 10^{-14}$ s$^{-1}$ for $z=0.03$
and $6.38 \times 10^{-14}$ s$^{-1}$ for $z=0.2$.}
\label{fig.PS_other_z}
\end{figure*}

To investigate whether this discrepancy is because of not considering uncertainty in 
the values of $\Gamma_{\rm HI}$ when we match $\rm dN/dz$ (see Fig.~\ref{fig.gama}), 
we estimate the 1-$\sigma$ lower and upper bounds on the $\Gamma_{\rm HI}$ by taking 
Poisson error on the measured $\rm dN/dz = 209.5 \pm 14.9$. Then by matching $\rm dN/dz$ 
we find $\Gamma_{\rm HI} = 7.73^{+0.87}_{-0.75} \times 10^{-14}$ s$^{-1}$ for IllustrisTNG. 
For this range of $\Gamma_{\rm HI}$ we estimate flux power and plot it as a cyan shade in the
Fig.~\ref{fig.ps}, which is still well below the measurements. 
Nevertheless, incorporating this uncertainty fails to resolve the discrepancy between 
the data and the models; instead, it shows that the discrepancy is quite severe. 

We further investigated whether the observed discrepancy could be attributed to the computation 
of power using perfect skewers, thereby neglecting the influence of real noise, window function 
corrections associated with the COS line spread function, and metal line masks used in power 
spectrum calculations from observational data \citep[as in][]{Khaire19}. To address this, we 
employed our forward models, consisting of $\approx 35000$ spectra, with metal masks 
positioned identically to those in the actual data. The power spectrum was then computed with 
appropriate window corrections, employing the Lomb-Scargle periodogram \citep{Lomb76, Scargle82} 
to handle the masked spectra, in accordance with the methods employed in 
\citep{Khaire19, Walther18a}, for both the Illustris and IllustrisTNG forward models. The 
results, depicted in Fig.~\ref{fig.ps} as red (Illustris) and blue circles (IllustrisTNG), 
reveal that these forward-modeled power spectra closely resemble the one obtained with 
perfect skewers. Thus, the observed discrepancy can not be attributed to the use 
of perfect versus forward-modeled spectra or the algorithms employed for power spectrum 
calculations.

To determine if the discrepancy between 
the measured power spectrum and the simulated power spectrum, 
for the values of $\Gamma_{\rm HI}$ obtained by matching the 
$\rm dN/dz$, exist at other redshifts, we conducted the same analysis at 
 two other redshifts  $z=0.03$ and $0.2$.
We generated forward-modeled skewers from simulation snapshots at $z=0.03$ and $z=0.1$ 
for different values of $\Gamma_{\rm HI}$ and used VPfit to fit these skewers and the 
\citet{Danforth14} data in the respective redshift bins. 
We then estimated $\Gamma_{\rm HI}$ by matching model $\rm dN/dz$ values at both redshifts with
$\rm dN/dz$ from observed data (see Fig.~\ref{fig.dndz_other_z} in the Appendix). 
For these $\rm dN/dz$ matched values of $\Gamma_{\rm HI}$, we computed the power spectra 
from perfect skewers and compared with our measurements from \citet{Khaire19}. 
The results are shown in Fig.~\ref{fig.PS_other_z}. Note that since the power spectrum from
forward-modeled \lya~forest matches closely with the one computed with perfect skewers 
(as shown in Fig.~\ref{fig.ps}), we do not show the forward-modeled power spectrum 
in Fig.~\ref{fig.PS_other_z}.
At $z=0.03$ and $0.2$, the simulated power spectra match well with the measured power spectra. 
There is no indication of the large discrepancy as seen for $z=0.1$. 
In fact, the reason behind the discrepancy seen here at  $z=0.1$ 
is because of non-monotonically increasing $\rm dN/dz$ with redshift. 
The values of $\rm dN/dz$ for  $z=0.03$ and $0.1$ and $0.2$ are
$227.5$, $205.9$ and $284.9$, respectively. The $\rm dN/dz$ decreases at $z=0.1$ as compared
to $z=0.03$ whereas higher values of $\rm dN/dz$ at $z=0.1$ would have reduced the discrepancy. 
All of these suggest that there may be an issue with the 
measured power spectrum or $\rm dN/dz$ at $z=0.1$. 
To investigate this we conducted various tests
and rechecked our calculations performed in \citet{Khaire19} with 
different data sets and processing of the data, but we were unable to resolve 
the problem. We have also noted a similar discrepancy in the case of NyX simulations.
This problem at $z=0.1$ warrants further investigation of 
the power spectrum measurement and perhaps $\rm dN/dz$ measurement in order to fully 
understand the nature of the disagreement between simulations and observations.

However, when comparing the power spectra at all three redshifts from Illustris and 
IllustrisTNG, we observe good agreement (within 5\%) at large scales ($k< 0.02$ s 
km$^{-1}$), but noticeable differences arise at smaller scales. This indicates that the 
small-scale power spectrum ($k>0.02$ s km$^{-1}$) is indeed sensitive to the feedback 
prescriptions used in the simulations. The reason behind the difference is 
perhaps because of the relatively strong feedback model employed by Illustris. 
Nevertheless, it is important to note that the 
magnitude of this difference is not significantly large, and at even smaller scales 
($k>0.05$ s km$^{-1}$), the noise power dominates the measurements. 
But a more glaring discrepancy is in the measurements and the predictions at $z=0.1$ 
power spectrum. To address this discrepancy and further explore the sensitivity of the 
power spectrum to feedback at small scales, 
we intend to conduct thorough investigations in future work.

It is also worth noting that the feedback introduces variations in the power 
spectrum at small scales ($k>0.02$ s km$^{-1}$) of around 4-5\%. 
This could potentially introduce additional 
degeneracies when attempting to measure the thermal state of the IGM at low redshifts. While the 
effect of feedback is expected to diminish at higher redshifts ($z>2$), it may still 
be necessary to consider the small percentage level differences in the power spectrum 
resulting from feedback when using the high-$z$ \lya~power spectrum to obtain precise 
cosmological parameters, as indicated by \citet{Chabanier20}.

We also show the power spectrum for the IllustrisTNG simulation using the \citet{FG09} UV 
background at three different redshifts: $z=0.1$, $z=0.03$, and $z=0.2$ in Fig.~\ref{fig.ps} 
and \ref{fig.PS_other_z}. In Fig.~\ref{fig.ps}, we 
verify the findings of \citet{Burkhart22} that the IllustrisTNG power spectrum at $z=0.1$ calculated for \citet{FG09} UV background is 
consistent with measurements. However, upon inspecting the power spectrum at the other redshifts 
(as shown in Fig.~\ref{fig.PS_other_z}), we observe that it is significantly higher than 
the measurements for 
most $k$ values. This implies that the match between the IllustrisTNG power spectrum and 
measurements at $z=0.1$ using the \citet{FG09} UV background may be merely coincidental.

\section{Discussion}\label{sec.disc}
In galaxy formation simulations, AGN feedback is necessary to regulate star formation 
in massive galaxies, which drives powerful outflows and heats the surrounding gas. 
However, the central AGN that provides this feedback cannot be resolved in these 
cosmological simulations. 
Therefore simulations adopt a sub-grid implementation of feedback. 
There are many different feedback prescriptions used in simulations, and their 
parameters are tested and adjusted using a range of observations. Most of these 
observations are based on the properties and distribution of galaxies and the gas very 
close to them (within $\sim R_{\rm vir}$). Unfortunately, there are no benchmark 
observations that extend beyond galaxies and into the IGM to 
provide constraints on these feedback prescriptions. This is particularly important 
because reproducing the IGM in simulations is relatively straightforward based on well-
established theory. Thus, if feedback can affect the IGM, it may be possible to use 
observations of the IGM itself to constrain the feedback prescriptions in simulations.

In this study, we sought to determine whether the IGM can be 
used to distinguish between different feedback prescriptions in simulations, using 
Illustris and IllustrisTNG as test cases. These simulations implement different 
feedback models, as described in Section~\ref{sec.sims_ill}. Illustris is known for its 
explosive feedback, which removes large amounts of gas from massive halos and heats it 
to high temperatures, making it a useful extreme case for our analysis. The 
temperature and density distribution in the simulations 
(shown in Fig.~\ref{fig.boxplot} and \ref{fig.T-Delta}) 
clearly demonstrate that the IGM is impacted by the feedback. Therefore
we studied if different statistics of the \lya~forest can be used to distinguish
the feedback used in Illustris and IllustrisTNG simulation so that those statistics
can be used for constraining the feedback prescriptions.  

In order to compare these two simulations with observations, 
we need to ensure that any potential degeneracies with observables are accounted for.
Since the optical depth of the Ly$\alpha$ forest depends on the product of
fraction of gas in diffuse \lya~phase and $\Gamma_{\rm HI}$ (see Eq.~\ref{eq.fgpa}), 
we first tuned $\Gamma_{\rm HI}$ to match the $\rm dN/dz$ with observations. 
Once we do that we find that it is very hard to detect any difference in the various statistics
such as joint or separate $b$-\nhi~distribution or \HI~CDD even when using the highest
S/N subset of HST COS \lya~forest data from \citet{Danforth14} in our forward models. 
However, only the \lya~flux power spectrum at small spatial 
scales appears to be sensitive to feedback.

Our conclusions are in disagreement with those of \citet{Burkhart22}, who also 
investigated the effects of feedback on the IGM using Illustris and IllustrisTNG 
simulations. They did not set the $\Gamma_{\rm HI}$ by matching ${\rm dN/dz}$ or 
any other statistics while comparing with observations,  but they did consider the degeneracy of  
$\Gamma_{\rm HI}$ with AGN feedback. 
Despite this, they concluded that the \lya~forest can serve as a 
diagnostic of AGN feedback. However, in our analysis, we have used 
comprehensive forward models using the observations, fitted the 
Voigt profiles to \lya~forest in both simulations 
and data using the same code, and conclusively 
shown that the small differences in various statistical measures 
are too hard to detect with current HST-COS data.

 It is also worth noting that 
\citet[][see their Appendix C]{Tepper-Garcia12} performed a similar
analysis presented here but with Overwhelmingly Large simulations \citep[OWLS;][]{Owls_sim}. Their findings revealed a negligible effect of feedback on the $z=0.5$ IGM even without tuning the $\Gamma_{\rm HI}$, consistent with the results of \citet{Theuns02} conducted at high-$z$. This is in contrast with what we find
in the Illustris simulation, perhaps because of the extreme feedback employed in this particular simulation.

It is important to note that since our work is focused on investigating if IGM 
observations can be used to probe feedback, we need to address the 
degeneracy of the \lya~forest with respect to the unknown UV background.
However, if the goal is to just compare the impact of feedback on the IGM in 
two simulations in order to build physical intuition, 
one can just use one unique value of $\Gamma_{\rm HI}$ in post-processing
or run both simulations with the same UV background to make the trends 
in various statistics easier to understand. 
The Illustris and IllustrisTNG are ideal sets of simulations where 
both are run with the same UV background 
model of \citet{FG09}.
In such cases where simulations are processed with the same UV background, 
all the \lya~forest statistics will 
show huge differences if the gas fractions in diffuse \lya~phase are significantly 
different. For example, Fig.~\ref{fig.gama} shows that
for the same $\Gamma_{\rm HI}$ the $\rm dN/dz$ of IllustrisTNG is a factor of 1.5-2 times
higher than Illustris. A similar magnitude effect can be seen in power spectrum 
\citep[see][]{Burkhart22} and CDD \citep[see][]{Gurvich17}.
In fact, not just Illistris and IllustrisTNG simulations but 
even Simba simulations with different feedback prescriptions show huge
differences in mean flux decrement \citep{Christiansen19} and CDD 
\citep{Tilman23}.

In \citet{Gurvich17}, it was argued that the combination of the \citet{FG09} UV 
background and AGN feedback prescription in Illustris is more realistic, as the CDD 
from Illustris matches well with the measurement from \citet{Danforth14}. However, 
this is not the only possible interpretation, as the same conclusion could be drawn 
for IllustrisTNG with the $\Gamma_{\rm HI}$ used here, i.e. the unknown 
$\Gamma_{\rm HI}$ degree of freedom was not varied in their analysis. 
In fact, a similar conclusion was 
reached by \citet{Tilman23}, who argued that the AGN jet feedback in the Simba 
simulation, along with its UV background from \citet{HM12}, is more realistic. This is 
in contrast to the conclusion of \citet{Christiansen19}, who found that the AGN jet 
feedback in the Simba simulation, along with the UV background of \citet{FG09}, is 
more realistic.\footnote{The difference in the conclusions of \citet{Christiansen19} 
and \citet{Tilman23} may have arisen due to the use of different statistics, with the 
former using flux decrement and the latter using the CDD.} In fact, as shown in 
section~\ref{sec.gamma}, the difference in $\Gamma_{\rm HI}$ in simulations
required to match observations depends on the diffuse \lya~fraction of the gas. 
Therefore, it is important to note 
that without independent constraints from UV background measurements or a direct
constraint on the gas fraction in the diffuse \lya~phase, 
it is not possible to definitively use the 
IGM to constrain the feedback model and perhaps metal lines \citep[see][]{Mallik23}
or clustering of \lya~absorbers \citep[see][]{Maitra22} might prove important.
However, note that
by fixing $\Gamma_{\rm HI}$ via ${\rm dN/dz}$ matching, the small scale
\lya~flux power spectrum appears to be sensitive to feedback. 

At present, the estimation of the UV background lacks simulation-independent 
constraints, as the measurements at low-redshift using \lya~forest
\citep[see][]{Gaikwad17a, Khaire19} 
rely on IGM simulations. Moreover, the simulations used in these studies do
not incorporate feedback from galaxy formation. However, 
simulation independent constraints on $\Gamma_{\rm HI}$
such as those obtained through quasar proximity zones \citep[][]{Bajtlik88} and 
through the analysis of 21 cm and H$\alpha$ emission lines in the outskirts of nearby 
galaxies \citep[][]{Dove94, Adams11, Fumagalli17}, would play a vital role to
study the effect of feedback on IGM. 
Nevertheless, these measurements are not as precise as the current 
measurements derived from the \lya~forest. Furthermore, 
there is no consensus on how much gas is in different phases either from observations
or simulations. Therefore, because of these issues, 
the value of $\Gamma_{\rm HI}$ must be treated as a  nuisance parameter while studying 
the impact of feedback on the IGM. 

In recent work, \citet{Tilman23} claimed that the low-redshift H~{\sc i} CDD
can be reproduced by the Simba simulation, 
providing evidence for the long-range jet AGN feedback implemented in that simulation. 
While their CDD does show good agreement with the measurements of \citet{Danforth14} 
compared to Illustris and IllustrisTNG simulations, there is still uncertainty in the 
role of the UV background in this agreement. The CDDs with and without jet feedback in 
the Simba simulation show similar slopes, with only small differences appearing at lower 
column densities, which is contrary to the differences seen  at higher column densities
(see Fig.~\ref{fig.cdd}) in the Illustris and IllustrisTNG simulations as a result of 
feedback. Therefore, it may not be definitive 
evidence for their feedback model. However, the different slopes of the CDD with and 
without jet feedback seen in the Simba simulation \citep[Figure 4 of][]{Tilman23} 
do suggest that Simba simulation appears to be
better at reproducing the CDD shape.

It is also worth noting that the studies by \citet{Gurvich17}, \citet{Christiansen19}
and \citet{Tilman23} were partly motivated by the discrepancy pointed out by
\citet{Kollmeier14}, who argued that their simulation needed a five times higher 
$\Gamma_{\rm HI}$ than \citet{HM12} to reproduce the CDD of \citet{Danforth14}. 
This higher required value of $\Gamma_{\rm HI}$ to match the Ly$\alpha$ forest 
CDD was thought to possibly be a problem for UV background synthesis models, 
but later \cite{Khaire15puc} showed that it is easy to obtain such high values with
updated UV background models. Other studies 
\citep[][]{Shull15, Gaikwad17a, Khaire19} even showed that simulations 
without any feedback require just a factor of two higher UV background, 
which is consistent with a UV background dominated by quasars 
alone \citep[][]{Khaire15puc, KS19, Khaire16} without needing any contribution of 
ionizing photons from galaxies. In fact, if the true feedback is as extreme as in Illustris,
leading to a reduction of gas fraction in diffuse \lya~phase as compared to no-feedback models, 
then the actual UV background intensity would be even lower than what is sourced 
by quasars alone. This scenario can provide interesting challenges for UV background 
synthesis models. 

\section{Summary}\label{sec.summary}

The objective of this study is to assess the influence of AGN feedback on the 
IGM and, consequently, explore the potential of the low-redshift 
\lya~forest as an effective means to probe feedback models.
For this analysis, we used two state-of-the-art simulation outputs, 
Illustris and IllustrisTNG at $z=0.1$. These simulations provide an ideal 
laboratory for understanding the impact of AGN feedback on the IGM due to their 
shared initial conditions and nearly identical cosmological parameters,
while exhibiting significant differences in AGN feedback strength and characteristics.
The main difference that can potentially affect the gas around the 
galaxies and in the IGM is the implementation of the AGN feedback in the radio mode.
Specifically, Illustris employs a bubble model in which a large amount of energy is 
accumulated before being released explosively, while IllustrisTNG employs a 
kinetic wind model in which momentum is imparted to nearby cells stochastically.
We sought to determine if the differences in AGN 
feedback could affect the IGM and, therefore, if observations of the IGM could be 
used to constrain AGN feedback models.

It is evident that the feedback is disrupting the gas around 
galaxies, forming large bubbles of hot gas that extend far into the IGM 
up to $10-20$ pMpc from the nodes of the cosmic web (Fig.~\ref{fig.boxplot}). 
This can be observed in the temperature-density phase diagram 
(Fig.~\ref{fig.T-Delta}), where the fraction of WHIM gas is 
significantly higher in Illustris (62\%) compared to IllustrisTNG (40\%), while the 
fraction of diffuse \lya~gas is significantly lower in Illustris (23\%) 
compared to IllustrisTNG (39\%).

To determine if the current data on low-redshift \lya~forest can be used
to probe these differences we generated synthetic \lya~forest spectra from both
simulations. Because the fraction of gas in the diffuse low-temperature phase
that is responsible for the observed \lya~absorption is quite different in 
both simulations, we tune the UV background in each simulation by adjusting 
$\Gamma_{\rm HI}$ in such a way that both simulations reproduce the observed line density ${\rm dN/dz}$. 
For this, we created forward models using the noise properties of observational data
\citep[][]{Danforth14} and fit both forward models and observations using the
same automatic Voigt profile fitting code (see Fig.~\ref{fig.gama}).
We found that to match our ${\rm dN/dz}$ measurement at $z=0.1$ IllustrisTNG requires 
$\Gamma_{\rm HI} = 7.73 \times 10^{-14} $ s$^{-1}$, whereas 
Illustris requires $\Gamma_{\rm HI} = 3.1 \times 10^{-14} $ s$^{-1}$.
The difference in the values of $\Gamma_{\rm HI}$ obtained for both simulations
are consistent with the fraction of diffuse \lya~gas in each simulation.

This tuning of $\Gamma_{\rm HI}$ is essential for accurately comparing simulations 
to real data. By demanding that the simulations match simple 
statistics of the observed IGM, such as the ${\rm dN/dz}$, we are bringing both 
simulations on an equal footing. 
Once we have fixed $\Gamma_{\rm HI}$ we investigate the \lya~forest statistics 
that can be used to probe the effect of feedback on the IGM. For that,
we studied various statistics, 
including the joint $b$-\nhi~distribution, the marginal $b$ and \nhi~distributions, 
H~{\sc i} CDD, and the flux power spectrum. 
The main results of these analyses are as follows. 

Using our Voigt profile decomposition of \lya~lines 
we study the $b-$\nhi~distributions of the IGM. For perfect high quality 
\lya~forest spectra (${\rm S/N} =100$ per pixel scale of $\Delta v = 4.1$ km 
s$^{-1}$), we find that the contours of $b-$\nhi~distributions 
clearly show differences between the two simulations (Fig.~\ref{fig.bN_igm_perfect}). 
Illustris shows more gas at high $b$ low \nhi~regions whereas IllustrisTNG has more 
gas at low $b$ high \nhi. These differences are mainly seen in the 
\lya~absorption at \nhi$<10^{12.5}$ cm$^{-2}$. 
However, when we forward model real HST COS data 
differences in the $b-$\nhi~distributions are washed out 
(Fig.~\ref{fig.bN_igm_perfect} $-$ \ref{fig.bN_igm_separate})
because the noise and finite resolution of the data limits its sensitivity to
lower \nhi~values. The HST-COS data can effectively probe \lya~lines only 
at \nhi$>10^{13}$ cm$^{-2}$ and therefore the forward modelled 
$b-$\nhi~distributions from both simulations become indistinguishable.  

We find that the \HI~CDD for both simulations (Fig.~\ref{fig.cdd})
is identical for \nhi$\lesssim 10^{14.5}$ cm$^{-2}$ for both forward and perfect
modeled \lya~forest. There is a small difference at
higher column densities  \nhi$>10^{14.5}$ cm$^{-2}$ for forward modelled spectra
and at \nhi$>10^{15}$ cm$^{-2}$ for perfect spectra.
Despite this, the magnitude of the difference is too small to be statistically 
significant based on the current HST-COS data.

We conducted a comparison of the \lya~forest flux power spectrum between 
the Illustris and IllustrisTNG simulations at three different redshifts: 
$z=0.03, 0.1$, and $0.2$. Our analysis reveals that the power spectra 
from both simulations exhibit consistency on large scales ($k<0.02$ s km$^{-1}$) 
with differences within 5\% between the two simulations. However, we observed 
a notable deviation at smaller scales, specifically in the range of 
$0.02<k <0.1$ s km$^{-1}$, where the Illustris simulation exhibited higher power. 
This finding suggests that the \lya~flux power spectrum  at small scales
is indeed sensitive to feedback, and further investigation is warranted to explore 
its potential, despite the dominance of noise power in this range of scales.

Based on our findings, it is clear that feedback has a non-negligible effect 
on the low-$z$ IGM, as is evident in the temperature density distribution of gas in 
Illustris and IllustrisTNG simulations (Fig.~\ref{fig.boxplot} and \ref{fig.T-Delta}).
When we examine various statistics of \lya~forests we find
a small but noticeable differences are seen 
in the full $b-$\nhi~distributions from \lya~forest 
in perfect spectra (S/N$=100$, resolution 4.1 km s$^{-1}$), 
on small scales in the line-of-sight \lya~flux power spectrum and 
at higher column densities in the H~{\sc i} CDD  
(\nhi$\, \gtrsim \, 10^{14.5}$ cm$^{-2}$). 
However, most of these differences are too small and the S/N and resolution of 
currently available data is not sufficient to definitively detect these 
effects except for the small-scale power spectra. 

This lack of sensitivity in various \lya~forest statistics arises because of 
the inherent degeneracy between the fraction of gas in diffuse \lya~phase 
(cool baryons) and the $\Gamma_{\rm HI}$ as their product determines the 
mean optical depth of the Ly$\alpha$ forest. 
Since the UV background cannot be precisely predicted from 
first principles, it needs to be treated as a nuisance 
parameter that we adjust by matching with the $\rm dN/dz$ from HST-COS data. 
Once we take this degree of freedom into account the potential 
distinctions in most statistics essentially fade away.
Only the Ly$\alpha$ flux power spectrum at small spatial scales 
($k > 0.02$ s km$^{-1}$) exhibits potentially observable distinctions, 
although this may be specific to the relatively extreme feedback model 
employed in Illustris. Nevertheless,
without independent constraints on either the UV background intensity 
or the fraction of gas in the diffuse \lya~phase, the task of constraining AGN 
feedback using the low-redshift Ly$\alpha$ forest remains challenging.

In the end, we want to bring attention to an intriguing issue with 
the \lya~power spectrum at $z=0.1$. 
We expected that the power spectra from our simulations would 
be consistent with observations when using the value of $\Gamma_{\rm HI}$ obtained by 
matching the $\rm dN/dz$ from the same observations. This proved to be true at $z=0.03$ 
and $z=0.2$ (as shown in Fig.~\ref{fig.PS_other_z}), but not at $z=0.1$ 
(see Fig.~\ref{fig.ps}). In this case, our simulated power spectrum is a 
factor of $1.5$ lower than the measurements on average. 
Despite conducting a thorough investigation, we were unable to identify the cause of 
this discrepancy. Further investigation is needed to understand the nature of 
this issue.

In our upcoming follow-up paper \citep[Paper II,][under review]{Khaire23}, we will 
explore the effect of feedback on the gas near the halos of massive galaxies by 
utilizing a range of \lya~forest statistics and available HST-COS data. 
Additionally, we will conduct a thorough investigation of the observed discrepancy in 
the \lya~flux power spectrum at $z=0.1$, employing different datasets and 
methodologies to identify the underlying cause of this discrepancy.

\section*{acknowledgement} 
VK thanks D. Sorini and D. Nelson for helping with Illustris simulation datasets.
VK thanks R. Srianand for hosting him at the Inter-University Centre for Astronomy 
and Astrophysics (IUCAA), Pune, India, and for helpful discussions on the paper. 
VK is supported through the INSPIRE Faculty Award (No.DST/INSPIRE/04/2019/001580) 
of the Department of Science and Technology (DST), India, and by NASA
through grant number HST-AR-17048.003 from the Space Telescope Science Institute,
which is operated by the Associated Universities for Research in Astronomy, Inc., 
under NASA contract NAS 5-26555.

\section*{Data Availability} 
We utilized a subset of quasar spectra from the dataset presented in 
\citet{Danforth14} which is accessible at 
\href{https://archive.stsci.edu/prepds/igm}
{https://archive.stsci.edu/prepds/igm}. The simulation data is obtained from the 
Illustris website, with links provided in the text at relevant locations.

\bibliographystyle{mnras}
\bibliography{vikrambib}

\appendix
\section{ The ${\rm dN/dz}$ matched $\Gamma_{\rm HI}$ at other redshifts}
\begin{figure*}
\begin{subfigure}{0.5\textwidth}
  \centering
  \includegraphics[width=.99\linewidth]{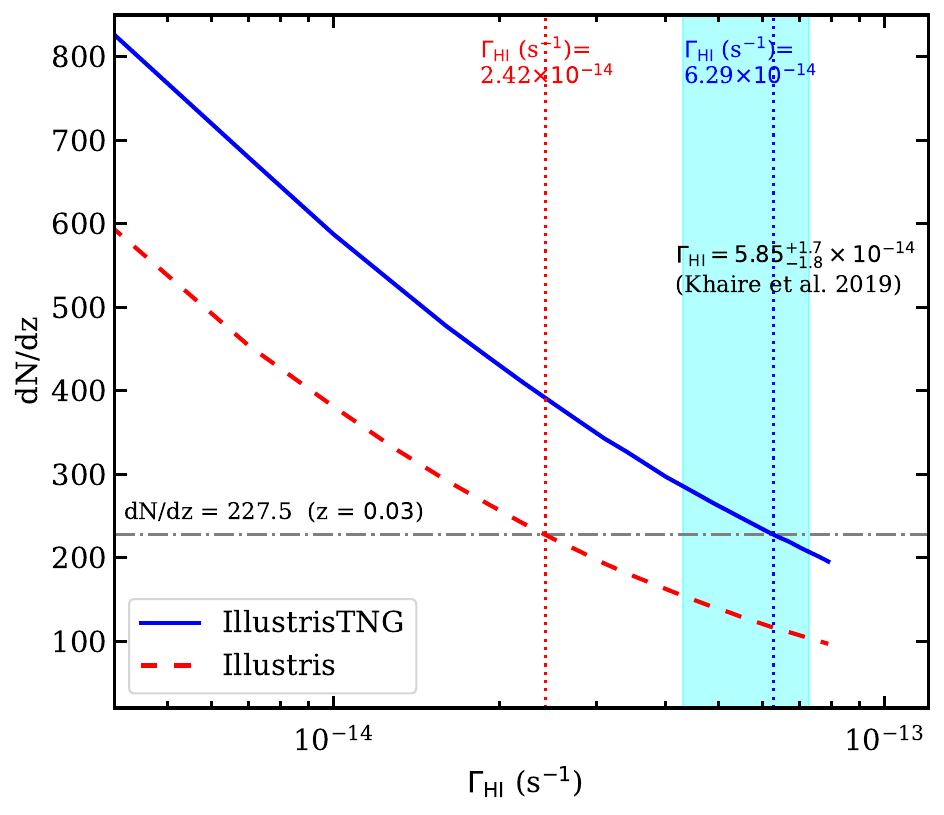}
  \label{fig:image1}
\end{subfigure}%
\begin{subfigure}{0.5\textwidth}
  \centering
  \includegraphics[width=.99\linewidth]{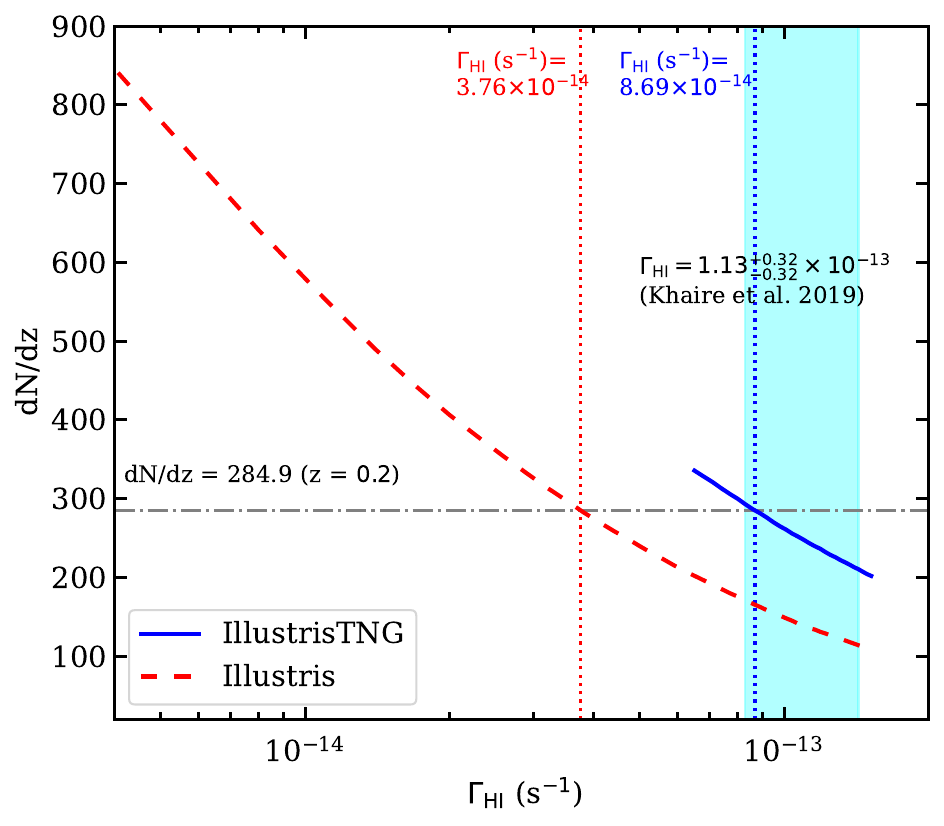}
  \label{fig:image2}
\end{subfigure}
\caption{ The line density ${\rm dN/dz}$ for \lya~absorbers in the range 
$10 ^{12} < N_{\rm HI} < 10^{14.5} \, { \rm cm^{-2} }$ is plotted for 
IllustrisTNG (blue curve) and Illustris (red dashed curve) simulations at
$z=0.03$ (left) and $z=0.2$ (right) using forward-modeled spectra. 
The grey dot-dashed line shows ${\rm dN/dz}$, obtained by fitting Voigt 
profiles to the \lya~forest observed in HST COS data from \citet{Danforth14} 
in the redshift bins $0.005<z<0.06$ (left-hand panel) and
$0.16<z<0.26$ (right-hand panel). The vertical dotted lines indicate the values of 
$\Gamma_{\rm HI}$ for Illustris (red) and IllustrisTNG (blue) at which the 
simulations match the ${\rm dN/dz}$ of the data. 
The cyan-shaded region shows the measurements of $\Gamma_{\rm HI}$ and 
its 1$\sigma$ uncertainty obtained by \citet{Khaire19}.
}
\label{fig.dndz_other_z}
\end{figure*}

Here, in Fig~\ref{fig.dndz_other_z} we show ${\rm dN/dz}$ as a function of $\Gamma_{\rm HI}$ 
for both Illustris and IllustrisTNG simulation at $z=0.03$ and $0.2$. These are used to find
the $\Gamma_{\rm HI}$ at these redshift for which the ${\rm dN/dz}$ from simulation match with the ${\rm dN/dz}$ measurements. 
The figure is similar to the Fig.~\ref{fig.gama} shown for $z=0.1$.

\end{document}